\documentclass{article}

\usepackage{PRIMEarxiv}

\usepackage[utf8]{inputenc} 
\usepackage[T1]{fontenc}    
\usepackage{hyperref}       
\usepackage{url}            
\usepackage{booktabs}       
\usepackage{amsfonts}       
\usepackage{nicefrac}       
\usepackage{microtype}      
\usepackage{lipsum}
\usepackage{fancyhdr}       
\usepackage{graphicx}       
\graphicspath{{media/}}     

\usepackage{amsmath,amssymb}
\usepackage{bm}
\newcommand{\mi}{{\rm i}}
\newcommand{\me}{{\rm e}}
\newcommand{\md}{{\rm d}}
\newcommand{\rr}{{\bm r}}
\newcommand{\Ai}{{\rm{Ai}}}
\newcommand{\Bi}{{\rm{Bi}}}

\title{Quantum Reflection Effects in the Diffraction of Matter Waves}

\author{
  Johannes Fiedler, Eivind K. Osestad \\
  Department of Physics and Technology\\ University of Bergen\\5007 Bergen, Norway\\
  \texttt{johannes.fiedler@uib.no} \\
   \And
  Fabian Spallek \\
Institut f\"ur Physik\\Universit\"at Kassel\\34132 Kassel, Germany\\
 \AND
  Andreu Vega \\
  Universitat de Barcelona\\08028 Barcelona, Spain\\
 \And
  Quentin Bouton, Gabriel Dutier \\
  Laboratoire de Physique des Lasers\\Université Sorbonne Paris Nord\\CNRS UMR 7538\\ 93430 Villetaneuse, France\\
}

\begin{document}
\maketitle

\begin{abstract}
    Among the fundamental quantum effects, quantum reflection (QR) is one of the most notable phenomena.
    Approximating arbitrary potentials in the Schr\"odinger equation as multistep potentials allows us to determine the reflection coefficient by means of a Riccati equation.
    We introduce a measure for the wavefunction's characteristic reflection distance and derive an analytical expression for this reflection point in an arbitrary power-law potential.
    We study the associated QR rate in the context of Casimir--Polder atom surface interactions, revealing its strong dependence on critical parameters such as velocity and interaction duration.
    As an application, we demonstrate how the matter-wave diffraction of a low-velocity metastable Argon beam impinging on a rotated grating facilitates high-precision potential analysis.
\end{abstract}


\section{Introduction}
Quantum reflection is a striking phenomenon in quantum mechanics where a particle reflects from a potential without encountering a classical barrier~\cite{Brenig1980,Boeheim1982,PhysRevA.65.032902,PhysRevLett.91.193202,10.1063/1.3246162,PhysRevLett.97.093201}. Unlike classical mechanics, where reflection requires a hard wall or energy barrier, quantum reflection arises purely from the wave nature of particles and the subtle interplay between their motion and the shape of the potential. This phenomenon typically occurs in smoothly varying or long-range potentials, such as those produced by van der Waals or Casimir–Polder interactions~\cite{PhysRevA.71.052901,Buhmann12a,Buhmann12b,D2CP03349F}. In such cases, the gradual change in the potential causes a mismatch in the particle’s wave function, leading to partial reflection, even when the particle’s energy exceeds the local potential. 

Quantum reflection plays a crucial role in various fields, ranging from atomic physics to surface physics. Thus, it has attracted significant interest as a tool for exploring the atom-surface potentials~\cite{RevModPhys.81.1051}, as well as for designing innovative quantum sensors~\cite{PhysRevA.108.023306,Fiedler2024} and new atomic optical devices such as mirrors and cavities~\cite{Oberst2005}. The first experimental studies on quantum reflections were conducted with incident atoms interacting with liquid helium~\cite{Ite1993} and a supersonic metastable atomic beam~\cite{Grucker2007}. Subsequently, this phenomenon was observed and investigated with cold beams at grazing incidence on both solid flat surfaces~\cite{PhysRevLett.86.987} or micro-structured surfaces~\cite{Zhao2008}. More recently, quantum reflection has been studied in experiments involving Bose-Einstein condensates at normal incidence, exploring various surface geometries~\cite{Pasquini2004, Pasquini2006}, as well as utilising laser-induced potentials~\cite{Marchant2016}. From a theoretical point of view, quantum reflection has been explored in different situations. For example, some works focus on the objects involved in the reflection from the surface, such as antimatter particles~\cite{Dufour2014}, solitons~\cite{Bai2020}, or ultracold molecules~\cite{Bai2019}. Other studies concentrate on the surface itself, with topics like rough surface~\cite{Rojas2020} or the engineering of surface designs for controllable quantum reflection~\cite{Kilianski2024}. In these studies, quantum reflection is typically described by simulating the time evolution of a wave packet by solving the Schrödinger equation numerically or using the semi-classical Wentzel--Kramers--Brillouin (WKB) approach. 

Here, in this work, we present both analytical and semi-analytical solutions describing the quantum reflection for different potential shapes. We derive the reflection probability, the reflection rate, and the reflection distance from the surface. Specifically, these results are applied to a quantum diffraction experiment involving slow metastable argon atoms interacting with a nanograting~\cite{Lecoffre2025}.

\section{Propagation of matter waves}
The propagation of a non-relativistic, spinless particle of mass $m$ with a kinetic energy $E$ within a potential landscape $U(\rr)$ is determined by the Schr\"odinger equation~\cite{Lekner2016}
\begin{equation}
    \left[-\frac{\hbar^2}{2m}\nabla^2+U(\rr)\right]\psi(\rr)=E\psi(\rr)\,, \label{eq:schroedinger}
\end{equation}
with the reduced Planck constant $\hbar$. Assuming that the potential only varies along the $z$ direction, $U=U(z)$. Hence, $\psi$ depends only on longitudinal ($z$) and transverse changes regarding the potential direction, $\psi=\psi(x,z)$. Due to the independence of potential on the transverse direction, the remaining linearity of the Schr\"odinger equation~(\ref{eq:schroedinger}) yields the assumption $\psi(x,z)=\me^{\mi K_x x}\Psi(z)$ leading to
\begin{equation}
    \frac{\md^2\Psi}{\md z}+q^2\Psi=0\,,\label{eq:schroedinger1d}
\end{equation}
with the longitudinal wave vector component
$q(z)=\sqrt{\frac{2m}{\hbar^2}\left[E-U(z)\right]-K_x^2}$.
By introducing the particle's de-Broglie wavelength $\lambda_{\rm dB} = h/p = h/\sqrt{2mE}=2\pi/K$, the wave number $K$ of the entire wave can be identified. Thus, the longitudinal wave vector is modified by the potential
\begin{equation}
    q(z)=\sqrt{K_z^2 -\frac{2m}{\hbar^2}U(z)}\,.
\end{equation}

\section{Reflection at a potential step}\label{sec:step}
The consideration of the reflection at a potential step is described by
\begin{equation}
U(z)=
\left\{\begin{array}{lr}
U_1&{\rm{for}}\, z < 0\\
U_2&{\rm{for}}\, z>0
\end{array}\right. \,,
\end{equation}
is an ordinary textbook example; see, for instance, Ref.~\cite {Lekner2016}. By inserting the step potential into Eq.~(\ref{eq:schroedinger1d}), the resulting solution in each region reads
\begin{equation}
    \Psi(z_{{}_<^>}) = C_{{}_<^>}^{(1)}\me^{\mi q_{1,2}z}+C_{{}_<^>}^{(2)}\me^{-\mi q_{1,2} z}\,,
\end{equation}
with
    $q_{1,2}(z)=\sqrt{K_z^2  - \frac{2m}{\hbar^2}U_{1,2}}$.
According to the continuity of  $\Psi(z)$ and $\Psi'(z)$ at $z=0$, the coefficients need to obey the relations
\begin{eqnarray}
    C_>^{(1)}&=&\frac{(C_<^{(1)}+C_<^{(2)})q_2+(C_<^{(1)}-C_<^{(2)})q_1}{2q_2}\,, \label{eq:D1}\\ 
    C_>^{(2)}&=&\frac{(C_<^{(1)}+C_<^{(2)})q_2-(C_<^{(1)}-C_<^{(2)})q_1}{2q_2}\,. \label{eq:D2}
\end{eqnarray}
By considering a wave hitting the interface from the left ($z<0$), the coefficients $C_<^{(1)}$ denote the incoming intensity, $C_<^{(2)}$ is the intensity of the reflected wave, and $C_>^{(1)}$ is the intensity of the transmitted wave. The remaining coefficient vanishes $C_>^{(2)}=0$ because it would depict an incoming wave from the right-hand side. This means that the reflection coefficient is the ratio between the reflected and incoming intensity $r=C_<^{(2)}/C_<^{(1)}$, and analogously, the transmission coefficient can be determined $t=C_>^{(1)}/C_<^{(1)}$. By inserting these conditions into Eq.~(\ref{eq:D2}), one directly obtains the reflection coefficient
    $r=\frac{q_1-q_2}{q_1+q_2}$,
which is equivalent to the reflection of electromagnetic $s$-waves at a dielectric step. Analogously, the transmission coefficient can be found $t=\frac{2q_1}{q_1+q_2}$. Reflectance and transmittance are given by the squares of the corresponding absolute values, $R=\left|r\right|^2$ and $T=\left|t\right|^2$, respectively. In analogy to the scattering of electromagnetic waves, these coefficients satisfy the relation $t=r+1$. In analogy to classical mechanics, the reflection properties depend on the particle's momentum, and due to the conservation of energy and momentum, the reflection at the interface follows Snell's law~\cite{Born1999,jackson1998classical}. The classical reflection properties of the particle are given when the perpendicular kinetic energy is smaller than the potential barrier. Otherwise, the particle transmits completely. In the quantum-mechanical picture, the total reflection of the particle at the interface is given when the wave vector in the right potential becomes imaginary $q_2\in\mathbb{C}$, which is satisfied by $U_2> \hbar^2K_z^2/(2m) = E-\hbar^2K^2/(2m)$, leading to an exponential decay of the wave function in this region, known as the evanescent field.

\section{Reflection at an arbitrary potential}
As previously demonstrated (in Sec.~\ref{sec:step}), matter waves behave like electromagnetic $s$-waves. Thus, the solution of the Schr\"odinger equation in a multistep potential
\begin{equation}
    U(z) = U_i \quad,\mathrm{if} \,z\in \left[z_i,z_{i+1}\right]\,,
\end{equation}
with the discontinuous boundaries located at $z_i$. Thus, based on the propagation of electromagnetic waves in planarly multilayered media~\cite{Chew}, the reflection of matter waves at the boundary at $z=z_i$ can be constructed from the elementary reflection and transmission coefficients 
\begin{equation}
r_{i,i+1} = \frac{q_{i+1}-q_i}{q_{i+1}+q_i} \,,\quad t =\frac{2q_i}{q_i+q_{i+1}}\,,\label{eq:refl}
\end{equation}
respectively, via the recursive formula
\begin{equation}
    \tilde{r}_{i,i+1} =  \frac{r_{i,i+1} + \tilde{r}_{i+1,i+2}\me^{2\mi q_{i+1} \left(z_{i+1}-z_i\right)} }{1 +r_{i,i+1}  \tilde{r}_{i+1,i+2}\me^{2\mi q_{i+1} \left(z_{i+1}-z_i\right)} }\,.
\end{equation}
Considering an equidistant layering $z_i =\Delta i$ with $i\in\mathbb{Z}$, expressing the reflection coefficient $r(z)$, the generalised reflection coefficient $\tilde{r}(z)$, and the wave vector $q_{i+1} =q(z-\Delta)$ at the boundary $z=-d_i$, the generalised reflection coefficient can be written in continuous variables
\begin{equation}
    \tilde{r}(z) =\frac{r(z) +\tilde{r}(z-\Delta) \me^{2\mi q(z-\Delta)\Delta}}{1 +r(z)\tilde{r}(z-\Delta)\me^{2\mi q(z-\Delta)\Delta}}\,.\label{eq:reflcontinuous}
\end{equation}
By applying the continuum limit ($\Delta\mapsto 0$), the reflection takes the form $r(z) = q'(z)/(2q(z))\Delta$, with $q(z) = \sqrt{K_z^2-(2m/\hbar^2)U(z)}$, the propagator and generalised reflection coefficient can be series expanded, leading to $\tilde{r}(z-\Delta) \approx \tilde{r}(z) -\Delta \tilde{r}'(z)$, $\me^{2\mi q(z-\Delta)\Delta}\approx 1+\Delta 2\mi q(z)$. Thus, the continuum limit of Eq.~(\ref{eq:reflcontinuous}) can be performed and yields
\begin{eqnarray}
\tilde{r}(z) = \tilde{r}(z) +\frac{q'(z)}{2q(z)}\Delta+2\mi q(z)\tilde{r}(z) \Delta -\tilde{r}'(z)\Delta -\frac{q'(z)}{2q(z)}\tilde{r}^2(z) \Delta \,.
\end{eqnarray}
Hence, the reflection coefficient is determined by the Riccati equation
\begin{equation}
    \tilde{r}'(z) = 2\mi q(z)\tilde{r}(z) +\frac{q'(z)}{2q(z)}\left[1-\tilde{r}^2(z)\right]\,. \label{eq:ricatti}
\end{equation}
This equation is usually solved numerically via the Runge--Kutta method in a forward scheme for a vanishing initial reflection coefficient $r(z\mapsto-\infty)=0$ in a potential region where the wave only transmits. Both solving the Schr\"odinger equation~\eqref{eq:schroedinger1d} for the wave and the Ricatti equation~\eqref{eq:ricatti} for the reflection coefficient for singular potentials yield issues concerning the boundary conditions at the interface. In a repulsive potential (due to Pauli repulsion), the Schr\"odinger and Ricatti equations will yield perfect reflection due to the infinite potential barrier. However, we know from experiments that several effects occur when scattering particles off a surface. Parts of the wave will scatter elastically (coherently) and parts will scatter inelastically, e.g. partial or complete transfer of kinetic energy; the latter will lead to physisorption, or even chemical reactions can occur, leading to chemisorption. Many different ways exist dealing with this issue, most of them are restricted to the consideration of coherently scattered particles; hence, absorption models need to be defined, which can be done, for instance, by introducing complex-valued potentials, absorption regions surrounding the bodies and forcing the wave function to be zero. As our main focus is the description of quantum reflection, we can consider the reflection to occur at a large distance from the surface. This allows us to define a certain reflection distance, where the potential is assumed to be constant and thus a well-defined part of the wave transmits further, and the remaining part is coherently reflected. Details on this consideration are given in the supplementary information.

\section{Position of reflection}
For quantum reflection in continuously varying potentials, no singular specific reflection point can be given due to the wave nature of the phenomenon. 
However, a useful quantitative substitute measure may be derived from analysing the validity of the Wentzel--Kramers--Brillouin (WKB) approximation.
Recall the WKB solutions as applicable to slowly varying potentials and particles with large kinetic energies as solutions to the (effectively) one-dimensional Schrödinger equation~(\ref{eq:schroedinger1d}) given by
\begin{equation}
\psi_{\rm WKB} (z) \approx \frac{1}{\sqrt{p(z)}}  \mathrm{exp}\left[\pm \frac{\mi}{\hbar} \int p(z')\md z' \right] \,,  
\end{equation}
with local classical momentum
    $p(z) =\sqrt{2m \left[E_\perp-U(z)\right]}$,
and with the particle's kinetic energy perpendicular to the surface $E_\perp = m v_\perp^2/2$.
Introducing 
\begin{equation}
    B(z) = \hbar^2\left(\frac{3}{4}\frac{\left[p'(z)\right]^2}{p^4(z)} - \frac{1}{2}\frac{p''(z)}{p^3(z)}\, ,\right)\label{eq:badlands_B}
\end{equation}
as proposed in~\cite{PhysRevA.65.032902}, the WKB wave functions provide good approximations to the exact solutions of the Schrödinger equation when $\left|B(z)\right|\ll 1 $. (Note that a more straightforward condition for the validity of the WKB approximation is discussed in~\cite{PhysRevLett.86.987}, but it is neither necessary nor sufficient.) Typically, this condition is maintained for smoothly varying potentials in regions of very small or very large $z$, where the potential changes gradually. 
However, $\left|B(z)\right|$ also immediately quantifies regions, so-called badlands, where the WKB approximation fails due to rapid variations in the potential.

Quantum reflection then occurs in these badlands, even in the absence of any explicit potential barrier. This reflects the phenomenon's inherently quantum-mechanical nature, which arises purely from the interplay between the particle's wavelike behaviour and the structure of the potential landscape. Finally, we may take $z_{\rm r}$ where the maximum of $\left|B(z)\right|$ occurs and $\left|B(z)\right|\ll 1 $ is maximally violated as a suitable point that characterises the reflection ~\cite{c9030064,doi:10.1126/sciadv.1500901,PhysRevA.94.012513,Doak_2000}. 

Since we are particularly interested in Casimir-Polder interactions, we consider specifically a potential of the form 
\begin{equation} \label{eq:PowerPotential}
    U(z)= \lambda z^n + U_0 
\end{equation}
for integer $n$ and real $\lambda$ and $U_0$, usually considered in the literature~\cite{PhysRevB.64.085418}. As shown in the supplementary information, the reflection occurs at
\begin{eqnarray}
    z_{\rm r} = \left(\frac{8-5n^2+\left(\sqrt{21n^2 +18n-39}-3\right)n}{n^2+6n+8}\frac{E_\perp-U_0}{\lambda}\right)^{\frac{1}{n}}\,,\nonumber\\\label{eq:z0}
\end{eqnarray}
or at $ z_{\rm r}=\left|\frac{n-2}{\frac{5}{2}n-4}\frac{E_\perp-U_0}{\lambda}\right|^{1/n}$ for potentials for exponents $n=-2$ and $n=-4$. Attractive potentials are described by negative prefactors $\lambda<0$, ensuring a real and positive solution for unbounded states, $E_\perp > U_0$.

\section{Quantum reflection of metastable argon}
Let's consider the reflection of meta-stable argon atoms with the polarisability $\alpha(\mi\xi)$ off a silicon nitride surface with the dielectric function $\varepsilon_{\rm P}(\mi\xi)$ at a temperature $T$. In this case, the interaction between the atoms and the surface is determined by the Casimir--Polder potential~\cite{Scheel2008,PhysRevB.101.235424,Osestad2025}
\begin{eqnarray}
U_{\rm CP}(z) = \frac{k_{\rm B}T \mu_0}{4\pi}\sum_{n=0}^\infty {}' \xi_n^2\alpha(\mi\xi_n) \int\limits_0^\infty \mathrm d k^\parallel \, \frac{k^\parallel}{\kappa^\perp} \mathrm e^{-2\kappa^\perp z}\left[r_s-\left(2\frac{{k^\parallel}^2 c^2}{\xi_n^2 }+1\right)r_p\right] ,\label{eq:UCP}
\end{eqnarray}
with the Boltzmann constant $k_{\rm B}$, the Matsubara-frequencies $\xi_n = 2\pi n k_{\rm B} T/\hbar$, the perpendicular wave vector
    $\kappa^\perp = \left( {k^\parallel}^2 + \frac{\xi^2}{c^2}\right)^{1/2}$,
and the Fresnel-reflection coefficients for \textit{s}- and \textit{p}-polarised waves
\begin{eqnarray}
  r_s = \frac{\kappa^\perp -\kappa_{\rm P}^\perp}{\kappa^\perp +\kappa_{\rm P}^\perp} \,,\quad r_p = \frac{\varepsilon_{\rm P}\kappa^\perp -\kappa^\perp_{\rm P}}{\varepsilon_{\rm P}\kappa^\perp +\kappa^\perp_{\rm P}}\,,
\end{eqnarray}
with $\kappa^\perp_{\rm P} =  \left( {k^\parallel}^2 + \varepsilon_{\rm P}(\mi\xi) \xi^2/c^2\right)^{1/2}$. The primed sum denotes that the first term has to be weighted by $1/2$. This quantum effect has been studied in several experiments~\cite{Schollkopf1994,https://doi.org/10.1002/andp.201500214,Hemmerich16} and theories~\cite{Pars,Fiedler15,C9CP03165K,Fiedler_2022}. The optical responses are taken from Refs.~\cite{PhysRevA.85.042513,PhysRevResearch.6.023165}. For short (\textit{non}-retarded limit) and large (\textit{ret}arded limit) atom-surface distances, the Casimir--Polder potential simplifies to the
\begin{equation}
    U_{\rm non}(z) = -\frac{C_3}{z^3} \, , \quad U_{\rm ret}(z) = -\frac{C_4}{z^4}\,,\label{eq:Uasy}
\end{equation}
respectively. A detailed discussion on these limits can be found in Refs.~\cite{PhysRevA.85.042513,PhysRevB.101.235424}. For the reflection of argon off a silicon nitride surface, the parameters $C_3 = 8.1296\cdot10^{-49}\,\rm{Jm^3}$ and $C_4 = 1.1778\cdot 10^{-55}\,\rm{Jm^4}$ can be obtained. Details are given in the supplementary information.

\begin{figure}[tb]
    \centering
    \includegraphics[width=0.6\linewidth]{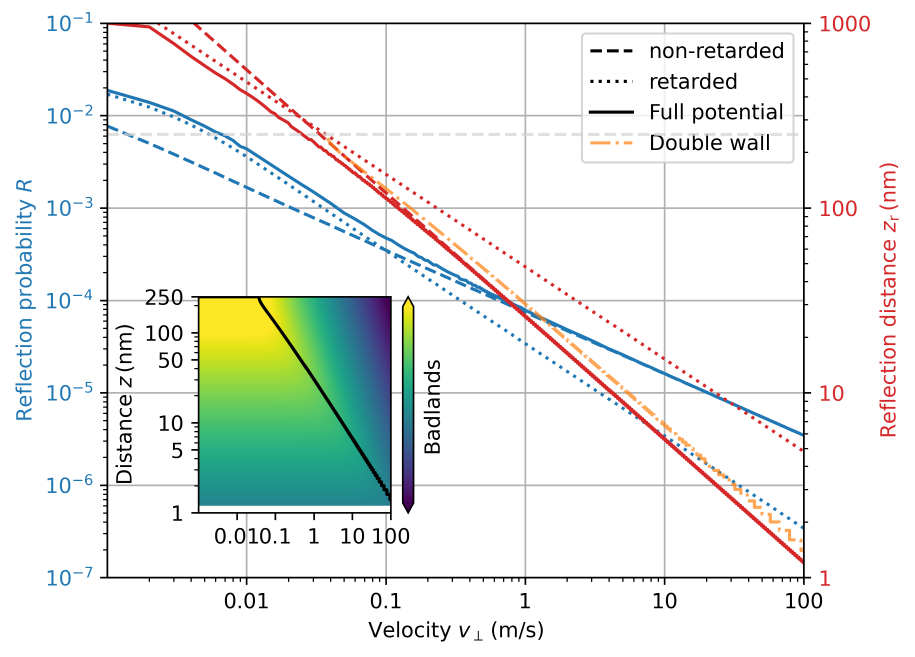}
    \caption{The reflection probability (blue lines) and reflection distance (red lines) of meta-stable argon reflected off a silicon nitride surface for different interaction potentials [non-retarded $C_3$-potential (dashed lines), retarded $C_4$-potential (dotted lines), full potential (solid lines)]. The orange line illustrates the breakdown of the reflection distance caused by the presence of an opposite surface at a distance of 500 nm, which occurs at a distance of 238 nm, illustrated by the dashed grey line. The inset depicts the badlands function (in arbitrary units) for the double-wall potential for different velocities. The black line shows the reflection distances as the maxima of the badlands for fixed velocities.} 
    \label{fig:reflectionlines}
\end{figure}

Inserting the parameters of the asymptotic solutions into (\ref{eq:z0}), one finds that the reflection occurs at
\begin{equation}
    z_{\rm r, non} = \sqrt[3]{\frac{C_3}{4E\left(3\sqrt{6}-7\right)}}\,,\quad     z_{\rm r, ret} = \sqrt[4]{\frac{C_4}{E}}\,,
\end{equation}
for the non-retarded potential and 
for the retarded potential, respectively. A particle being reflected at the Casimir--Polder potential comes from $+\infty$, reaching the returning point $z_0$ close to the surface before disappearing at $+\infty$ again. Thus, the particle interacts with the full potential and not only with its asymptotes. To incorporate this fact and to stay with a simplified potential, the potential
\begin{eqnarray}
    U(z) = -\frac{C_4}{z^3\left(z+\frac{C_4}{C_3}\right)}\,,\label{eq:C3C4}
\end{eqnarray}
is often used, satisfying both limits. Figure~\ref{fig:reflectionlines} depicts the reflection points depending on the velocity of the incoming beam for different approximations of the interaction potential. The intermediate potential~(\ref{eq:C3C4}) shows similar results to the full potential~(\ref{eq:UCP}). Analogously to the potentials, the asymptotic forms~(\ref{eq:Uasy}) will approximate the exact results in their corresponding regimes determined by the reflection distances. For large perpendicular velocities, the particle deeply penetrates the potential and reaches the non-retarded regime, which is well approximated by a full non-retarded potential. Analogously for small velocities, the returning point will be in the retarded regime, and consequently, an approximation of the reflection probability with a pure $C_4$-potential will be sufficient. As the gradient and curvature of the potential are the determining quantities of the reflection distance, the impact of an opposite surface, by considering the impact of a neighbouring grating bar as part of the following section, is dramatic. Adding the potential of a second surface at a distance of 500 nm yields a breakdown of the reflection distance at 238 nm, which an argon atom can reach with a perpendicular velocity of 0.038 m/s. Hence, slower particles will not be reflected as a result of the interaction with the opposite grating bar. The resulting reflection coefficient, obtained by numerical integration of the Ricatti differential equation~(\ref{eq:ricatti}) for an argon atom in the Casimir--Polder potential~(\ref{eq:UCP}) with the initial condition $r(z\mapsto -\infty) = 0$ and integrated until the returning point, see Fig.~\ref{fig:reflectionlines}. It can be observed that the reflection probability dramatically drops for high beam velocities. Fast particles reaching closer to the surface are well described by the non-retarded interaction potential,  whereas slower particles, which are reflected further away from the surface, are well approximated by the retarded potential. Similarly to the reflection point, the intermediate potential yields very similar results compared to the full potential.

\begin{figure}[t]
    \centering
    \includegraphics[width=0.6\linewidth]{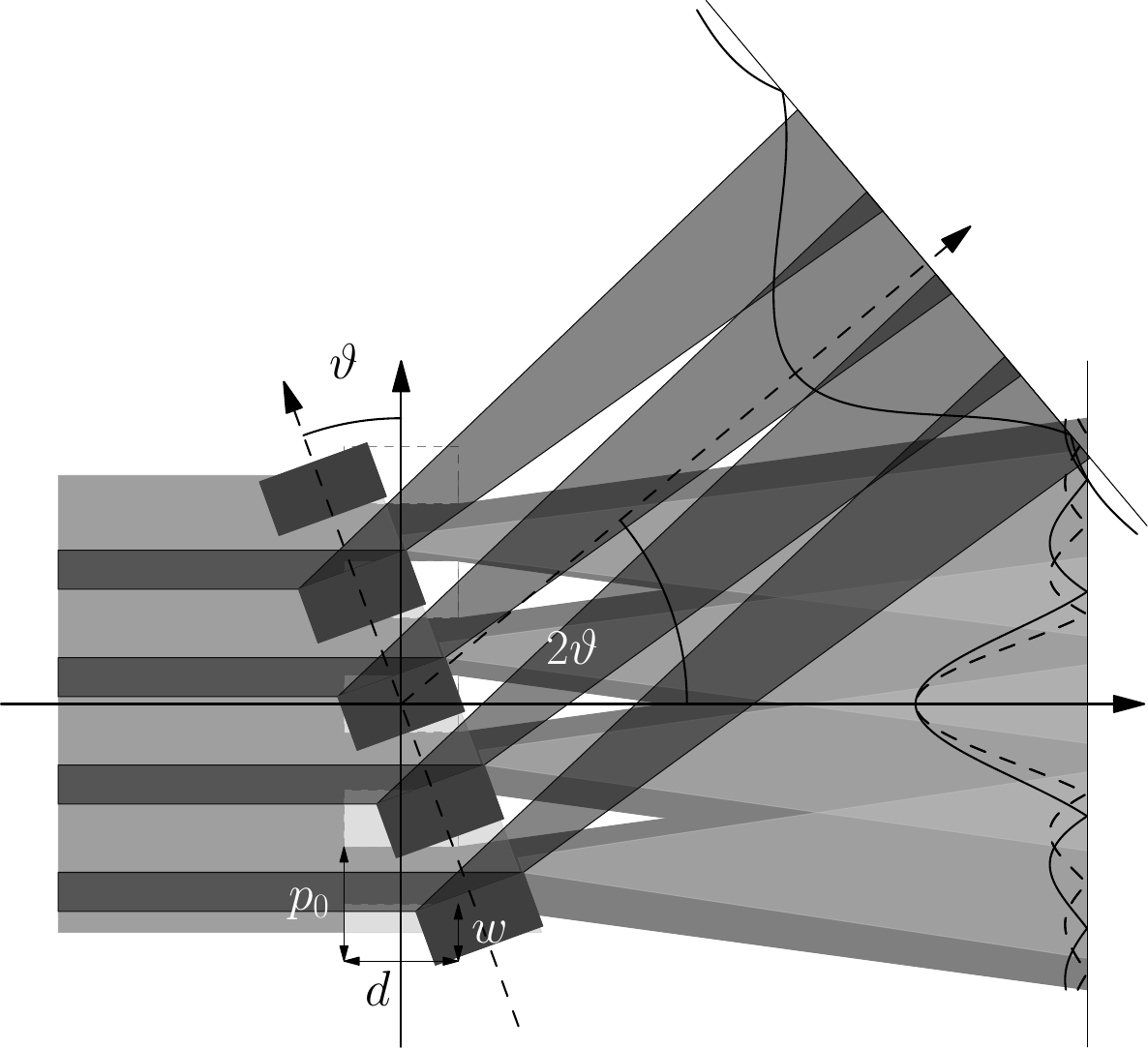}
    \caption{Schematic sketch of a matter-wave diffraction experiment with a rotated diffraction grating with width $w$, thickness $d$ and period $p_0$. The effective period decreases by rotating the grating by an angle $\vartheta$. In addition to this projected decrease, the effective opening is further reduced by the inward motion of the sidewall, providing reflective surfaces. Thus, a second interference pattern is expected at twice the rotation angle.}
    \label{fig:rotatedgrating}
\end{figure}

\section{Quantum reflection in matter-wave diffraction experiments}
Matter-wave diffraction experiments typically operate in transmission, e.g. through a dielectric grating~\cite{Arndt1999,Reisinger11,PhysRevLett.125.050401,Fiedler_2023}. For simplicity, let's consider a periodic grating with perfectly rectangular grating bars with the thickness $d$, the width $w$ and the period $p_0$. Hence, the wave will be blocked along the distance $w$ and pass through the grating along the distance $p_0-w$. By rotating the grating by an angle $\vartheta$, see Fig.~\ref{fig:rotatedgrating}, the sidewalls will be rotated into the beam axis, and the particles will be partially reflected, reaching this area. By rotating the grating, the period will decrease with the cosine of the rotation angle
\begin{equation}
    p(\vartheta) = p_0\cos\vartheta \,,\label{eq:period}
\end{equation}
according to the projection into the transverse plane. Due to the rotation of the grating, the area of the grating opening splits into two regions: a reflective region, which is determined by the inwards-moving sidewall, with a reflective length (rectangular area with infinite length)
\begin{equation}
    a(\vartheta) = d\sin\vartheta \,,\label{eq:reflarea}
\end{equation}
and a transmission region, which is given by the subtraction of the projected grating opening and the reflective region
\begin{equation}
    o(\vartheta) = (p_0-w)\cos\vartheta - d \sin\vartheta \,.
\end{equation}
Thus, the maximum rotation angle is determined by $o(\vartheta_{\rm max}) = 0$ and reads $\vartheta_{\rm max} = \arctan (p_0-w)/d$. The inward-moving sidewall defines the number of reflection atoms. Thus, a rotated grating will macroscopically split the beam with a transmission rate
\begin{equation}
T(\vartheta) = \frac{o(\vartheta)}{p(\vartheta)} = \frac{p_0-w}{p_0} -\frac{d}{p_0}\tan\vartheta\,,\label{eq:transrate}
\end{equation}
and reflection rate
\begin{equation}
    R(\vartheta) = r(K\sin\vartheta) \frac{d}{p}\tan\vartheta \,,\label{eq:reflrot}
\end{equation}
\begin{figure}[t]
    \centering
    \includegraphics[width=0.7\linewidth]{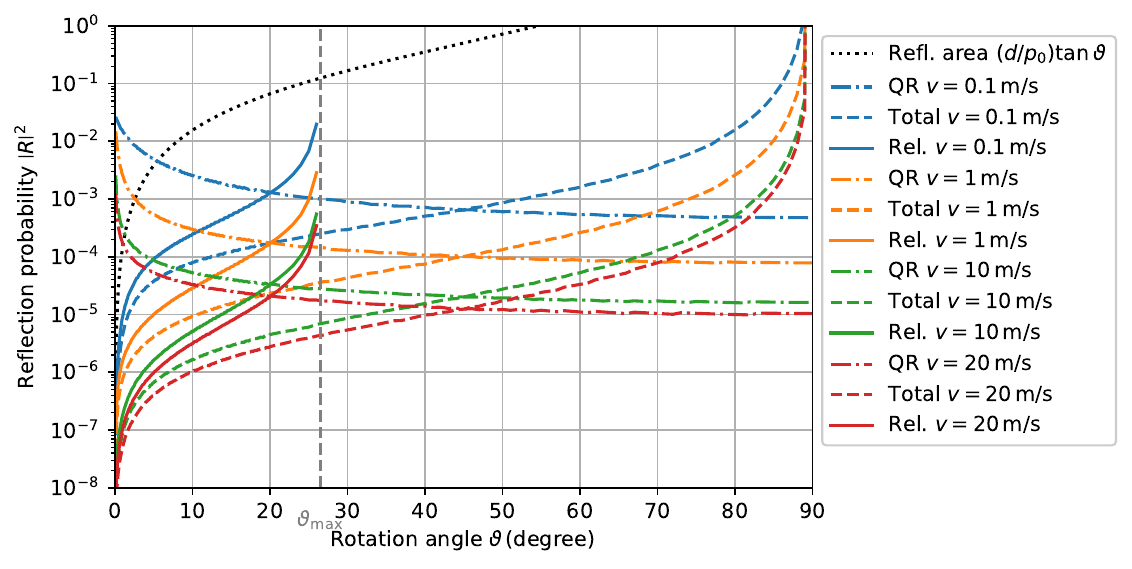}
    \caption{Plot of the reflection rate~(\ref{eq:reflrot}) for different particle velocities [0.1 m/s (blue curves), 1 m/s (orange curves), 10 m/s (green curves), 20 m/s (red curves)]. The reflection coefficients due to quantum reflection (labelled with QR) are depicted by the dashed-dotted lines. The black dotted line shows the increase in the reflection area~\eqref{eq:reflarea}. The product of both yields the reflection rate drawn by the coloured dashed lines. Relative reflection rates as the ratio between the reflected beam~\eqref{eq:reflrot} and the transmitted beam~\eqref{eq:transrate} are depicted by the solid lines.}
    \label{fig:reflrot}
\end{figure}
as the product of the reflection coefficient $r(K\sin\vartheta)$ with the wave vector of the matter wave $K$ due to quantum reflection, obtained by solving the Ricatti differential equation~\eqref{eq:ricatti} evaluated at the reflection point determined by the Badlands function~\eqref{eq:badlands_B} and the reflective area~\eqref{eq:reflarea}. Figure~\ref{fig:reflrot} illustrates the reflection rate for different velocities of an argon beam reflected off a silicon nitride grating with period $p_0=1000\,\rm{nm}$, width of the grating bar $w=500\,\rm{nm}$ and thickness $d=1000\,\rm{nm}$. The reflection coefficients for the quantum reflection are illustrated by dashed-dotted lines. The black dotted line indicates the geometric increase in the reflecting area. The product of both yields the reflection rates, depicted with solid lines. It can be seen that the geometric effect results in an enhancement of the reflection probability, whereas the quantum reflection probability decreases with increasing rotation angle. This reflection rate only describes the coherently scattered atoms diffracted via quantum reflection. Hence, both rates will not sum up to 1, $R+T<1$. We neglect reflection from the inner repulsive potential because metastable particles will most likely decay during scattering. 

Considering far-field diffraction (large distances between the source and the grating and the grating and the screen) allows us to describe the emergence of the interference patterns via the Kirchhoff diffraction formula~\cite{Born1999}. The diffraction of matter waves at transmission gratings has been discussed extensively in recent decades. So, we will restrict ourselves to describing the reflective part, which will be diffracted under twice the rotation angle. As sketched in Fig.~\ref{fig:rotatedgrating}, the interference along the reflected beam can be described via the Huygens--Fresnel principle as elementary waves generated at the tilted sidewalls. Hence, it can be treated as a periodic grating as well, with the same period as the transmission~\eqref{eq:period} and an opening area given by $d\sin\vartheta /p_0$. Due to the periodicity of the grating, the reduction of the atom number by coherent quantum reflection will affect the entire interference pattern in total. Finally, the remaining bit to compute the interference patterns is the phase accumulated by bypassing the grating. For matter-wave diffraction at dielectric objects, it is well-known that the particles will experience a spatially dependent phase shift due to the Casimir--Polder interaction. By considering a small rotation angle, which is required for high reflectivities due to quantum reflection as described above, the impact of the neighbouring grating bar can be neglected. By further neglecting boundary effects, the side wall can be considered to be infinitely thick so that the interaction potential will solely depend on the perpendicular direction. The consideration of edge effects will be part of future studies. In this case, the phase shift caused by the propagation through the potential equals at all positions because all reflected trajectories are translationally invariant. Hence, this phase cancels on the screen, and only the geometric phase due to the rotation of the grating needs to be considered. Finally, the interference patterns for the reflected beam can be described by
\begin{eqnarray}
   I_{\rm R}(\alpha) \approx I_0R^2(\vartheta,K)\frac{\sin^2\left[N\omega_{\rm int}\left(\alpha-2\vartheta\right)\right]}{\sin^2\left[\omega_{\rm int}\left(\alpha-2\vartheta\right)\right]}
    \frac{\sin^2\left[\omega_{\rm diff}\left(\alpha-2\vartheta\right)\right]}{\left(\alpha-2\vartheta\right)^2}\,,
\end{eqnarray}
separating into a diffraction part with period $\omega_{\rm diff} = K d\sin\vartheta/2$ and an interference part with period $\omega_{\rm int} = Kp_0\cos\vartheta/2$. Thus, the interference pattern of the reflected wave reduces to spatially shifted patterns (by the double incidence angle) of the well-known electromagnetic patterns. Details on interference patterns are given in the supplementary information. One can observe that the envelopes of the interference patterns (both in transmission as well as in reflection) are determined by the opening of the gratings and thus scale with the sine of the rotating angle $\sin\vartheta$. In contrast, the diffraction parts of the interference patterns are determined by the grating's period, and thus scale with the cosine. This yields that the expected interference fringes will not be observable for small rotation angles due to their small spatial separation which will be below the resolution of existing measurement techniques, and consequently, only the amount of reflected and transmitted particles will be experimentally accessible, leading to the expected ratios between the reflected and transmitted particle numbers as depicted by the solid lines in Fig.~\ref{fig:reflrot}.

\section{Conclusion}
The paper aims to answer the non-trivial question of the quantum reflection's distance and coefficient of a wave function crossing a power law potential. Such a fundamental quantum mechanical phenomenon finds general expressions relevant in many situations. As an example, we choose an extensively well-studied textbook and experimental physics known as the Casimir--Polder interaction between a free-flying atom and a material surface. As it exhibits a different potential's power law over the distance, our general solution is especially adapted and can be derived for a transmission grating crossed by a velocity-controlled atomic beam. We find drastic differences between transmitted and reflected diffraction pictures. The most valuable are the sharp QR rate and the diffraction envelope over the grating's rotated angle with regard to the atomic beam axis. In addition to their strong parameter dependence, the QR rate and diffraction envelope follow only the rotation angle at a given potential power law. This appears as an ideal setup for further theoretical and experimental investigations, considering that a single easily controllable parameter - the rotation angle - gives rise to an extremely sensitive signal, which is itself highly linked to the potential. Then, the potential unicity for each rotating angle makes the technique even more efficient to probe the potential over the distance to the surface and whatever the power law, which is expected to vary from $n=-3$ to $n=-4$ and back to $n=-3$ at a very large distance. Such a meticulous investigation would emphasise subtle effects that are still only predicted theoretically~\cite{PhysRevA.109.032824,PhysRevA.97.023806,Hu:2004zu}.

\appendix

\section{Casimir--Polder potential}
To consider the reflection of meta-stable argon atoms with the polarisability $\alpha(\mi\xi)$ off a silicon nitride surface with the dielectric function $\varepsilon_{\rm P}(\mi\xi)$ at a temperature $T$, the interaction between the atoms and the surface is determined by the Casimir--Polder potential~\cite{Scheel2008,PhysRevB.101.235424}
\begin{eqnarray}
U_{\rm CP}(z) = \frac{k_{\rm B}T \mu_0}{4\pi}\sum_{n=0}^\infty {}' \xi_n^2\alpha(\mi\xi_n) \int\limits_0^\infty \mathrm d k^\parallel \, \frac{k^\parallel}{\kappa^\perp} \mathrm e^{-2\kappa^\perp z}\left[r_s-\left(2\frac{{k^\parallel}^2 c^2}{\xi_n^2 }+1\right)r_p\right] ,
\end{eqnarray}
with the Boltzmann constant $k_{\rm B}$, the Matsubara-frequencies $\xi_n = 2\pi n k_{\rm B} T/\hbar$, the perpendicular wave vector
    $\kappa^\perp = \left( {k^\parallel}^2 + \frac{\xi^2}{c^2}\right)^{1/2}$,
and the Fresnel-reflection coefficients for \textit{s}- and \textit{p}-polarised waves
\begin{eqnarray}
  r_s = \frac{\kappa^\perp -\kappa_{\rm P}^\perp}{\kappa^\perp +\kappa_{\rm P}^\perp} \,,\quad r_p = \frac{\varepsilon_{\rm P}\kappa^\perp -\kappa^\perp_{\rm P}}{\varepsilon_{\rm P}\kappa^\perp +\kappa^\perp_{\rm P}}\,,
\end{eqnarray}
with $\kappa^\perp_{\rm P} =  \left( {k^\parallel}^2 + \varepsilon_{\rm P}(\mi\xi) \xi^2/c^2\right)^{1/2}$. The optical responses are taken from Refs.~\cite{PhysRevA.85.042513,PhysRevResearch.6.023165}. By considering distances smaller than the particle's relevant dipole transition $z\ll c/\omega_{\rm max}$, the Casimir--Polder potential can be simplified to
\begin{equation}
    U_{\rm CP}(z) = -\frac{C_3}{z^3} \, ,
\end{equation}
with the well-known $C_3$-coefficient
\begin{equation}
    C_3 =\frac{k_{\rm B}T}{8\pi\varepsilon_0} \sum_{n=0}^\infty {}' \alpha(\mi\xi_n)\frac{\varepsilon_{\rm P}(\mi\xi_n)-1}{\varepsilon_{\rm P}(\mi\xi_n)+1} \, .\label{eq:C3}
\end{equation}
Analogously, one finds the retarded limit for $z\gg c/\omega_{\rm min}$ leading to
\begin{equation}
    U_{\rm CP}(z) = -\frac{C_4}{z^4} \, ,
\end{equation}
with the $C_4$-coefficient
\begin{eqnarray}
 C_4 = \frac{3\hbar c\alpha(0)}{64\pi^2 \varepsilon_0 } \int\limits_1^\infty \mathrm d v \, \left[\left(\frac{2}{v^2} -\frac{1}{v^4}\right)\frac{\varepsilon_{\rm P}v -\sqrt{v^2-1+\varepsilon_{\rm P}}}{\varepsilon_{\rm P}v +\sqrt{v^2-1+\varepsilon_{\rm P}}}-\frac{1}{v^4}\frac{v-\sqrt{v^2-1+\varepsilon_{\rm P}}}{v+\sqrt{v^2-1+\varepsilon_{\rm P}}}\right] \, .
\end{eqnarray}

\section{Reflection at a potential step}\label{sec:step2}
The consideration of the reflection at a potential step is described by
\begin{equation}
U(z)=
\left\{\begin{array}{lr}
U_1&{\rm{for}}\, z < 0\\
U_2&{\rm{for}}\, z>0
\end{array}\right. \,,
\end{equation}
is a well-known textbook example; see, for instance, Ref.~\cite {Lekner2016}. By inserting the step potential into the Schr\"odinger Equation, the resulting solution in each region reads
\begin{equation}
    \Psi(z_{{}_<^>}) = C_1\me^{\mi q_{1,2}z}+C_2\me^{-\mi q_{1,2} z}\,,
\end{equation}
with
\begin{equation}
    q_{1,2}(z)=\sqrt{K_z^2  - \frac{2m}{\hbar^2}U_{1,2}}\,.
\end{equation}
According to the continuity of  $\Psi(z)$ and $\Psi'(z)$ at $z=0$ the complete solution reads as
\begin{equation}
    \Psi(z) = 
    \left\{\begin{array}{lr}
    C_1\me^{\mi q_{1}z}+C_2\me^{-\mi q_{1} z} &{\rm{for}}\, z<0\\
    D_1\me^{\mi q_{2}z}+D_2\me^{-\mi q_{2} z} &{\rm{for}}\, z>0
    \end{array}\right.\,,
\end{equation}
with
\begin{eqnarray}
    D_1&=&\frac{(C_1+C_2)q_2+(C_1-C_2)q_1}{2q_2}\,, \label{eq:D11}\\ 
    D_2&=&\frac{(C_1+C_2)q_2-(C_1-C_2)q_1}{2q_2}\,. \label{eq:D21}
\end{eqnarray}
By considering a wave hitting the interface from the left ($z<0$), the coefficients $C_1$ denote the incoming intensity, $C_2$ is the intensity of the reflected wave, and $D_1$ is the intensity of the transmitted wave. The remaining coefficient vanishes, $D_2=0$, because it would depict an incoming wave from the right-hand side. This means that the reflection coefficient is the ratio between the reflected and incoming intensity $r=C_2/C_1$, and analogously, the transmission coefficient can be determined as $t=D_1/C_1$. By inserting these conditions into Eq.~(\ref{eq:D21}), one directly obtains the reflection coefficient
\begin{equation}
    r=\frac{q_1-q_2}{q_1+q_2}\,,
\end{equation}
which is equivalent to the reflection of electromagnetic $s$-waves at a dielectric step. Analogously, the transmission coefficient can be found
\begin{equation}
    t=\frac{2q_1}{q_1+q_2}\,.
\end{equation}
The reflectance and transmittance are given by the squares of the corresponding absolute values, $R=\left|r\right|^2$ and $T=\left|t\right|^2$, respectively. In analogy to the scattering of electromagnetic waves, these coefficients satisfy the relation
\begin{figure}[htb]
    \centering
    \includegraphics[width=0.7\columnwidth]{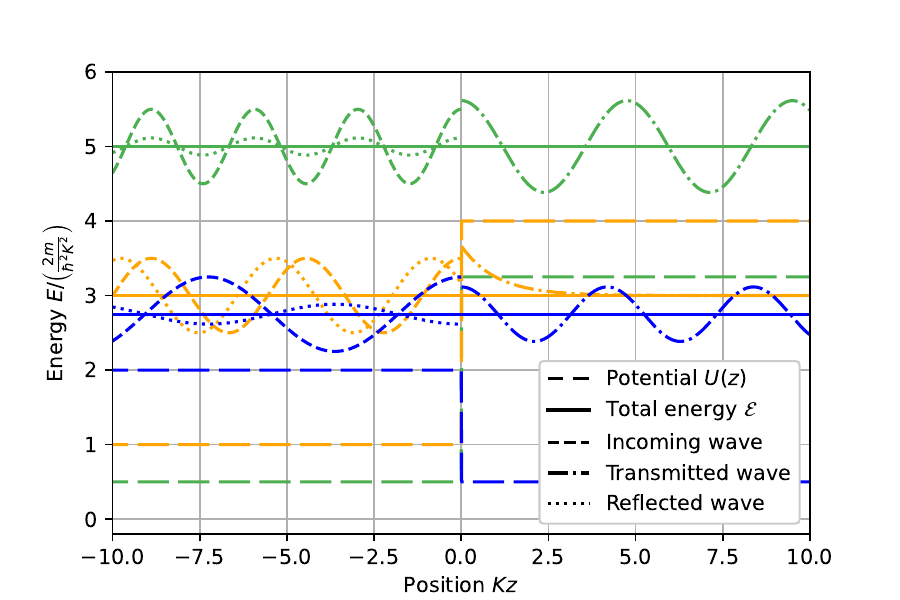}
    \caption{Reflection of a particle at a potential step with (green) and without at a rising (orange) and falling step (blue) transmission. For all cases, the potential $U(z)$ is depicted as long-dashed lines, the total energy $\mathcal{E}=\frac{2mE}{\hbar^2K^2}$ as solid horizontal lines, the incoming fields (short-dashed lines) as waves with the corresponding reflections (dotted lines) and transmissions (dashed-dotted lines).}
    \label{fig:step_ex}
\end{figure}
In analogy to classical mechanics, the reflection properties depend on the particle's momentum, and due to the conservation of energy and momentum, the reflection at the interface follows Snell's law~\cite{Born1999,jackson1998classical}. The classical reflection properties of the particle are given when the perpendicular kinetic energy is smaller than the potential barrier. Otherwise, the particle transmits completely. In the quantum-mechanical picture, the total reflection of the particle at the interface is given when the wave vector in the right potential turns imaginary, $q_2\in\mathbb{C}$, which is satisfied by $U_2> \hbar^2K_z^2/(2m) = E-\hbar^2K^2/(2m)$, leading to an exponential decay of the wave function in this region, known as evanescent field.  Figure~\ref{fig:step_ex} illustrates the different scenarios: (i) and (ii) the diffraction at a potential barrier (with falling potential step blue lines and rising potential step green lines), where the kinetic energy of the particle is larger than the potential step on both sides of the barrier leading to propagating waves with different wavelengths on both sides, according to the reflection law in classical optics; (iii) the reflection of a wave with partial absorption of the energy (orange lines) where the potential step is higher than the kinetic energy, in this case, the reflected wave is slightly amplitude-modulated and the transmitted part creates an evanescent field inside the right potential. 

\section{Special solutions for the reflection at spatially
continuous potentials.}
We consider the Schrödinger equation
\begin{equation}
    \Big[-\frac{\hbar^2}{2m}\nabla^2+U(\rr)\Big]\psi(\rr)=E\psi(\rr)\,,
\end{equation}
with a one-dimensional potential $U=U(z)$. Thus, we can separate equation via $\psi(\rr_\parallel,z)=e^{iK r_{\parallel]}}\psi(z)$ leading to
\begin{equation*}
    \frac{\md^2\psi(z)}{\md z^2}+\tilde{q}^2\psi(z)=0 \,,
\end{equation*}
with
\begin{equation*}
    \tilde{q}(z)=\sqrt{\frac{2m}{\hbar^2}[E-U(z)]-K^2}.
\end{equation*}
If we write
\begin{equation}
U(z) = 
\begin{cases} 
U_- & \text{for } z < 0, \\
\tilde{\Gamma}_n z^n + U_0 & \text{for } z \geq 0.
\end{cases}
\end{equation}
For $z<0$, assuming there is no incoming wave, we can surely write $\psi_-=C\me^{-\mi q_-z}$. For $z\geq0$ we need to solve
\begin{equation}\label{simplified_schrodinger_equation}
    \frac{\md^2\psi_n(z)}{\md z^2}+q^2\psi_n(z)+\Gamma_nz^n\psi_n(z)=0
\end{equation}
where
\begin{equation}
    q=\sqrt{\frac{2m}{\hbar^2}(E-U_0)},\ \ \ \ \text{and}\ \ \ \ \Gamma_n=-\frac{2m\tilde{\Gamma}_n}{\hbar^2}.
\end{equation}

\subsection{Reflection coefficient}
\begin{figure}
    \centering
    \includegraphics[width=0.4\linewidth]{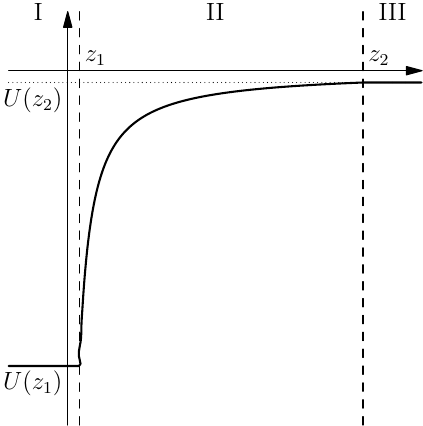}
    \caption{Fitting of the wave functions}
    \label{fig:fitting}
\end{figure}
Figure~\ref{fig:fitting} illustrates the fitting of the wave function to extract the reflection coefficient. The potential will be piecewise defined in three layers, composed of two constant potentials in regions I and III and a spatially varying potential in region II. The boundaries are according to Fig.~\ref{fig:fitting} at $z=z_1$ and $z=z_2$. Hence, the total potential reads 
\begin{equation}
    U(z)  = \begin{cases}
       U_2 = U(z_2) \, & \text{if} \,z> z_2 \\
       U_1 = U(z_1) \, & \text{if} \,z< z_1 \\
        U(z) \, & \text{else}
    \end{cases}\,.
\end{equation}
The wave arrives from $+\infty$, hence, the wave in region III is given by
\begin{equation}
    \psi_{\rm III} = C_1 \me^{-\mi q_2 z} + C_2 \me^{\mi q_2z}\,,
\end{equation}
with the amplitude of the incoming $C_1$ and of the reflected beam $C_2$. The transmitted beam is described by an outgoing wave in region I
\begin{equation}
\psi_{\rm I} = D_1 \me^{-\mi q_1 z}   \,.
\end{equation}
Thus, the reflection coefficient is defined by
$r= C_2/C_1$ and the transmission coefficient by $t=D_1/C_1$. Note that both depend on the positions $z_1$ and $z_2$. For the rational potentials ($1/r^n$), the limits $z_1\mapsto \infty$ should stay finite
\begin{eqnarray}
    R(z_1) = \lim_{z_2\mapsto \infty} r(z_1,z_2) \,,\\
    T(z_1) = \lim_{z_2\mapsto \infty} t(z_1,z_2) \,.
\end{eqnarray}
Finally, $z_1$ will be described via the Badlands functions.

\subsection{Non-singular potentials}
\subsubsection{Wave functions}
For $n=2$, the solution can be expressed as a combination of Whittaker
functions,
\begin{equation}\label{n=2}
    \Psi_2(z) = \frac{C_1}{\sqrt{z}} M_{-\frac{\mi q^2}{4 \sqrt{\Gamma_2}},\frac{1}{4}} \left( \mi \sqrt{\Gamma_2} z^2 \right)+ \frac{C_2}{\sqrt{z}} W_{-\frac{\mi q^2}{4 \sqrt{\Gamma_2}},\frac{1}{4}} \left( \mi \sqrt{\Gamma_2} z^2 \right)
\end{equation}

For \( n = 1 \), we obtain Airy functions,
\begin{equation}\label{n=1}
    \Psi_1(z) = C_1 \Ai \left( -\frac{q^2 + \Gamma_1 z}{\Gamma_1^{2/3}} \right) + C_2 \Bi \left( -\frac{q^2 + \Gamma_1 z}{\Gamma_1^{2/3}} \right).
\end{equation}

For $n=0$ it is a trivial combination of plane waves, and for $n=-1$ we can use Whittaker functions again,
\begin{equation}\label{n=-1}
    \Psi_{-1}(z) = C_1 M_{-\frac{\mi\Gamma_{-1}}{2q},\frac{1}{2}} \left( 2\mi qz \right) 
+ C_2 W_{-\frac{\mi \Gamma_{-1}}{2q},\frac{1}{2}} \left( 2\mi qz \right).
\end{equation}

In all these cases, the Hamiltonian is a self-adjoint operator. This means that imposing boundary conditions on the solutions of the Schrödinger equation leads to the quantisation of the energy spectrum, as well as a spectrum bounded from below, and the existence of ground states. Things start to get complicated from here on.

\subsubsection{Reflection and transmission coefficients}
Let us write $M(z)=M_{-\frac{\mi\Gamma_{-1}}{2q},\frac{1}{2}} \left( 2\mi qz \right)$ and $W(z)=W_{-\frac{\mi \Gamma_{-1}}{2q},\frac{1}{2}} \left( 2\mi qz \right)$ so that $\Psi_{-1}=AM(z)+BW(z)$. Fitting this solution to the step potential at $z=z_1$ gives the equations
\begin{eqnarray*}
    AM(z_1)+BW(z_1)=De^{\mi q_1z_1}
\end{eqnarray*}
and
\begin{eqnarray*}
    AM'(z_1)+BW'(z_1)=-\mi q_1D\me^{-\mi q_1z_1}\,,
\end{eqnarray*}
for $q_1=\sqrt{2m(E-U_1)/\hbar^2}$. This gives
\begin{eqnarray*}
    AM'(z_1)+BW'(z_1)=-\mi q_1[AM(z_1)+BW(z_1)] \,,
\end{eqnarray*}
so that
\begin{eqnarray}
    B=-\frac{M'(z_1)+\mi q_{\rm I}M(z_1)}{W'(z_1)+\mi q_{\rm I}W(z_1)}A=-C_{-1} A.
\end{eqnarray}
On the other hand, fitting to the step potential at $z=z_1$ gives the equations
\begin{eqnarray*}
    AM(z_2)+BW(z_2)=C_1\me^{-\mi q_2z_1}+C_2\me^{\mi q_2z_1}\,,
\end{eqnarray*}
and
\begin{eqnarray*}
    \frac{\mi}{q_2}[AM'(z_2)+BW'(z_2)]=C_1\me^{-\mi q_2z_2}-C_2\me^{\mi q_2z_2}\,.
\end{eqnarray*}
Adding and subtracting, we can easily find expressions for $C_1$ and $C_2$. Dividing them, we find
\begin{eqnarray*}
    \frac{C_2}{C_1}=\frac{\left(M(z_2)+\frac{\mi}{q_2}M'(z_2)\right)A+\left(W(z_2)+\frac{\mi}{q_2}W'(z_2)\right)B}{\left(M(z_2)-\frac{\mi}{q_2}M'(z_2)\right)A+\left(W(z_2)-\frac{\mi}{q_2}W'(z_2)\right)B}\me^{2\mi q_2z_2} \,.
\end{eqnarray*}
One now inserts $B(A)$ and takes the limit $z_2\rightarrow+\infty$, 
\begin{eqnarray*}
    \begin{aligned}
M_{-\frac{\mi\Gamma_{-1}}{2q}, \frac{1}{2}}(2\mi q z_1) &\sim \frac{\me^{\mi q z_1} (2\mi q z_1)^{-\frac{\mi\Gamma_{-1}}{2q}}}{\Gamma\left(1 - \frac{\mi\Gamma_{-1}}{2q}\right)}, \\
W_{-\frac{\mi\Gamma_{-1}}{2q}, \frac{1}{2}}(2\mi q z_1) &\sim \me^{-\mi q z_1} (2\mi q z_1)^{\frac{\mi\Gamma_{-1}}{2q}}, \\
M'(z_1) &\sim \mi q M(z_1), \\
W'(z_1) &\sim -\mi q W(z_1).
\end{aligned}    
\end{eqnarray*}
The reflectance reads
\begin{eqnarray*}
    R_{-1} \sim \left| \frac{\left(1 - \frac{q}{q_2}\right) - C_{-1} \left(1 + \frac{q}{q_2}\right) \Gamma\left(1 - \frac{\mi\Gamma_{-1}}{2q}\right)\me^{-2\mi qz_2}}{\left(1 + \frac{q}{q_2}\right) - C_{-1} \left(1 - \frac{q}{q_2}\right) \Gamma\left(1 - \frac{\mi\Gamma_{-1}}{2q}\right)\me^{-2\mi qz_2}} \right|^2
\end{eqnarray*}
where
\begin{eqnarray}
    C_{-1}=\frac{M'(z_1)+\mi q_1M(z_1)}{W'(z_1)+\mi q_1W(z_1)}.
\end{eqnarray}
Using $q_2=q$, this equation simplifies to
\begin{eqnarray}
    R_{-1}=\frac{\me^{-2\pi\frac{\Gamma_{-1}}{2q}}}{\left|C_{-1}\Gamma\left(1+\mi\frac{\Gamma_{-1}}{2q}\right)\right|^2}=\frac{e^{-2\frac{\pi\Gamma_{-1}}{2q}}}{\left|C_{-1}\right|^2}\frac{\sinh\frac{\pi\Gamma_{-1}}{2q}}{\frac{\pi\Gamma_{-1}}{2q}}\ .
\end{eqnarray}

\subsection{Inverse squared potential: anomaly and regularisation}
First of all, let us note an interesting feature of this potential: if we take \( U_0 = 0 \) (or simply absorb the term into the energy), the power-law form of the corresponding potential matches the order of the kinetic term. This results in a scale invariance of the solutions. Indeed, if \( \psi_{-2}(z) \) is a solution of the less manipulated Schrödinger equation
\begin{equation*}
    -\frac{\hbar^2}{2m}\frac{\md^2\psi_{-2}(z)}{\md z^2}-\frac{\tilde{\Gamma}_{-2}}{z^2}\psi_{-2}(z)=E\psi_{-2}(z)
\end{equation*}
then, for any $\beta\in\mathbb{R}$, we write $\psi_\beta(z)=\psi_{-2}(\beta z)$ and construct a new solution for
\begin{equation}
    -\frac{\hbar^2}{2m}\frac{\md^2\psi_{\beta}(z)}{\md z^2}+\frac{\tilde{\Gamma}_{-2}}{z^2}\psi_{-2}(z)=\beta^2 E\psi_{\beta}(z).
\end{equation}
This means that the existence of a single state with negative energy enables the construction of infinitely many eigenfunctions with eigenvalues as negative as desired. In other words, the system would lack a fundamental state. One can note, following Ref.~\cite{ovdat2019breaking}, that, in fact, any Hamiltonian
\begin{equation}
    H_N=\left(-\frac{\md^2}{\md z^2}\right)^N-\frac{\lambda_N}{z^{2N}},\ \ \ \ N\in\mathbb{Z},\ \lambda_N\in\mathbb{R},
\end{equation}
is subject to the same strange feature, referred to by the Lifshitz scaling symmetry. This is just a particular case of the so-called dimensional transumations, discussed more generally in Refs.~\cite{brattan2018landscape, camblong2001dimensional}.

A reasonable solution for that problem in $n=-2$ would be if there were no negative energy eigenstates. A trivial way to see when this is the case is to follow Refs.~\cite{essin2006quantum, gupta1993renormalization}, noticing that we can decompose
\begin{equation*}
    -\frac{\hbar^2}{2m}\frac{\md^2}{\md z^2}+\frac{\tilde{\Gamma}_{-2}}{z^2}=-\frac{\hbar^2}{2m}\left(\frac{\md}{\md z}+\frac{\nu}{z}\right)\left(\frac{\md}{\md z}-\frac{\nu}{z}\right)
\end{equation*}
for
\begin{equation}\label{index_a}
    \Gamma_{-2}=\nu(\nu-1)\implies\nu=\frac{1}{2}\pm\sqrt{1/4-a}
\end{equation}
and therefore
\begin{equation*}
    E=\langle\psi_{-2}|H|\psi_{-2}\rangle=-\frac{\hbar^2}{2m}\bigg\langle\psi_{-2}\bigg|\left(\frac{\md}{\md z}+\frac{\nu}{z}\right)\left(\frac{\md}{\md z}-\frac{\nu}{z}\right)\bigg|\psi_{-2}\bigg\rangle.
\end{equation*}
If $\nu$ is real, then 
\begin{equation*}
    E=\frac{\hbar^2}{2m}\int_0^{\infty}\bigg|\left(\frac{\md\psi_{-2}}{\md z}-\frac{\nu}{z}\right)\bigg|^2\md z>0
\end{equation*}
and there are no negative-energy eigenstates, avoiding the anomaly. However, if $\nu$ is not real, the problem must be confronted. From Eq.~(\ref{index_a}), this happens when $\Gamma_{-2}>1/4$, that is, 
\begin{equation}
    \tilde{\Gamma}_{-2}<-\frac{\hbar^2}{8m}\,.    
\end{equation}

Let us now find the general solution for $n=-2$ by using the Fröbenius method and see where these qualitative differences arise. The method used below is not necessary to solve this particular case of the Schrödinger equation; however, the change of variable moving the singularity to $\infty$ will allow us to compare the behaviour of the solution with the ones regarding other singular potentials (which need, indeed, to be solved by similar variable changes). Consider
\begin{equation}
    \psi''(z)+q^2\psi(z)+\frac{\Gamma_{-2}}{z^2}\psi(z)=0
\end{equation}
and change $y=1/z$, $\md/\md z=(\md/\md y)(\md y/\md z)=-y^2(\md/\md y)$, $(\md/\md z)^2=\md/\md z(-y^2\cdot \md/\md y)=-y^2(\md/\md y)(-y^2\cdot \md/\md y)=y^4(\md/\md y)^2+2y^3(\md/\md y)$. The previous equation now reads
\begin{equation*}
    \psi''(y)+\frac{2}{y}\psi'(y)+\Big(\frac{\Gamma_{-2}}{y^2}+\frac{q^2}{y^4}\Big)\psi(y)=0\,. 
\end{equation*}

 By expressing the solution as a descending power series, we obtain
\begin{equation*}
    \sum_{n=0}^{\infty}c_n(r-n)(r-n-1)y^{-n-2}+2\sum_{n=0}^{\infty}c_n(r-n)y^{-n-2}+\Gamma_{-2}\sum_{n=0}^{\infty}c_ny^{-n-2}+q^2\sum_{n=0}^{\infty}c_ny^{-n-4}=0
\end{equation*}
and we can write
\begin{equation*}
    \sum_{n=0}^{\infty}c_n\Big[(r-n)(r-n+1)+\Gamma_{-2}\Big]y^{-n-2}+q^2\sum_{n=0}^{\infty}c_ny^{-n-4}=0.
\end{equation*}
For $y^{-2}$, we get the equation
\begin{equation*}
    c_0[r(r+1)+\Gamma_{-2}]\rightarrow c_0=0\ \ \text{or}\ \ r=-1/2\pm\sqrt{1/4-\Gamma_{-2}}\,.
\end{equation*}
Set $c_0=0$ and see that for $y^{-3}$, we get
\begin{equation*}
    c_1[(r-1)r+\Gamma_{-2}]=0\rightarrow c_1=0\ \ \text{or}\ \ r=1/2\pm\sqrt{1/4-\Gamma_{-2}}\,.
\end{equation*}
Let us now choose the latter and give a solution in terms of $c_1$. One sees that this choice sets to zero all pair coefficients and, for odd coefficients, yields the expression
\begin{equation*}
    c_{2n+1}=(-1)^nq^{2n}c_1\prod_{i=0}^n\frac{1}{(r-2i-1)(r-2i)+\Gamma_{-2}}
\end{equation*}
and since we have chosen $r$ such that $r^2-r+\Gamma_{-2}=0$ this simplifies to
\begin{equation}
    c_{2n+1}=(-1)^nq^{2n}\frac{c_1}{4^n}\prod_{i=0}^n\frac{1}{i(i-r+1/2)}=(-1)^nq^{2n}\frac{c_1}{4^n}\prod_{i=0}^n\frac{1}{i(i+\tilde{a})}
\end{equation}
where we have defined $\tilde{a}=\mp\sqrt{1/4-\Gamma_{-2}}$. Note now that since the Pochhammer symbols are defined such that $(z)_n=z(z+1)\dots(z+n-1)=\Gamma(n+z)/\Gamma(z)$, we can write
\begin{equation}
    c_{2n+1}=(-1)^nq^{2n}\frac{c_1}{4^n}\frac{1}{n!\cdot (\tilde{a})_{n+1}}
\end{equation}
and one can write the solution in terms of
\begin{equation*}
    \psi(y)=c_1y^{r-1}\sum_{n=0}^{\infty}(-1)^n\frac{1}{n!\cdot (\tilde{a})_{n+1}}\Big(\frac{q^2}{4y^2}\Big)^n
\end{equation*}
and re-arranging
\begin{equation*}
    \psi(z)=\frac{c_1}{\tilde{a}}z^{1-r}\sum_{n=0}^{\infty}\frac{(-1)^n}{n!\cdot(\tilde{a}+1)_n}\Big(\frac{z\cdot q}{2}\Big)^{2n}=\frac{c_1\Gamma(\tilde{a}+1)}{\tilde{a}}z^{1/2\mp\sqrt{1/4-a}}\sum_{n=0}^{\infty}\frac{(-1)^n}{n!\cdot\Gamma(n+\tilde{a}+1)}\Big(\frac{z\cdot q}{2}\Big)^{2n}.
\end{equation*}
Now, recall that a Bessel function has the form
\begin{equation*}
    J_{\nu}(z)=\sum_{n=0}^{\infty}\frac{(-1)^n}{n!\cdot\Gamma(n+\nu+1)}\Big(\frac{z}{2}\Big)^{2n+\nu}.
\end{equation*}
Plugging $C_1=\frac{c_1\Gamma(\tilde{a}+1)}{\tilde{a}}$ and rearranging a bit, one sees that we can write
\begin{equation}
    \psi(z)=C_1\sqrt{z}\cdot J_{\tilde{a}}(q\cdot z).
\end{equation}
For the next step, recall that for non integer $\nu$, the functions $J_{\nu}(z)$ and $J_{-\nu}(z)$ are linearly independent. Since we have fixed the choice of $\tilde{a}$, a general solution is given by
\begin{equation}\label{solution_bessel}
    \psi_{-2}(z)=C_1\sqrt{z}\cdot J_{\sqrt{1/4-\Gamma_{-2}}}(qz)+C_2\sqrt{z}\cdot J_{-\sqrt{1/4-\Gamma_{-2}}}(qz).
\end{equation}

Since 
\begin{equation*}
    Y_\nu(z) = \frac{J_\nu(z) \cos(\nu \pi) - J_{-\nu}(z)}{\sin(\nu \pi)},
\end{equation*}
it is clear that one can redefine constants such that
\begin{equation*}
    \psi_{-2}(z)=C_1\sqrt{z}\cdot J_{\sqrt{1/4-\Gamma_{-2}}}(q z)+C_2\sqrt{z}\cdot Y_{\sqrt{1/4-\Gamma_{-2}}}(q z)\ .
\end{equation*}

Let us study closely the behaviour of this solution when defined over $[0,+\infty)$. According to \cite{case1950singular}, when the indexes of the Bessel functions in (\ref{solution_bessel}) are imaginary, both functions in the combination behave essentially the same. When studying bound states, one can express the solution in terms of a modified Bessel function of the second kind, $\psi_{-2}(z)=C\sqrt{z}K_{\sqrt{1/4-a}}(qz)$. The solutions are indeed square integrable
\begin{equation}
    \int_0^{\infty}\sqrt{z}K_{\sqrt{1/4-\Gamma_{-2}}}(qz)=\frac{\mi\pi\sqrt{1/4-\Gamma_{-2}}}{2q^2\sinh(\mi\pi\sqrt{1/4-\Gamma_{-2}})}\in\mathbb{R}
\end{equation}
so that the normalisation constant is
\begin{equation}
    C=q\sqrt{\frac{2\sinh(\mi\pi\sqrt{1/4-\Gamma_{-2}})}{\mi\pi\sqrt{1/4-\Gamma_{-2}}}}\in\mathbb{R}.
\end{equation}

To discuss the scattering for positive energy states, $E>0$, it is convenient to express (\ref{solution_bessel}) in terms of Hankel functions of the first two kinds, 
\begin{equation}
    \psi_{-2}(z)=\sqrt{z}[A\cdot H^{(1)}_{\sqrt{1/4-a}}(kz)+B\cdot H^{(2)}_{\sqrt{1/4-a}}(kz)],\ \ \ \ k=\sqrt{2mE/\hbar^2}
\end{equation}
defined as
\begin{equation}
    H_\nu^{(1)}(z) = J_\nu(z) + \mi Y_\nu(z)\ \ \ \ \text{and}\ \ \ \ H_\nu^{(2)}(z) = J_\nu(z) - \mi Y_\nu(z).
\end{equation}
For $z\rightarrow+\infty$ \cite{gradshteyn1988tables},
\begin{equation*}
    \psi_{-2}(z) \approx A \sqrt{\frac{2}{\pi k}} \me^{-\mi\frac{\pi}{2} (\sqrt{1/4-a}+1/2)} \left[ \me^{-\mi kz} - \mi \frac{B}{A} \me^{\mi\pi\sqrt{1/4-a}} \me^{\mi kz} \right]\,,
\end{equation*}
which, after some algebra, reduces to
\begin{equation}
    \psi_{-2}(z)\approx A \sqrt{\frac{2}{\pi k}} \me^{-\frac{\pi}{2} (\sqrt{a-1/4}+\mi/2)} \left[ \me^{-\mi kz} +\frac{B}{A} \me^{-\frac{3\pi}{2}kz} \me^{\mi\frac{3\pi^2}{2}\sqrt{a-1/4}} \right]\,,
\end{equation}
where the first term stands for the incident wave and the second stands for the reflected wave, decaying exponentially as expected.

It is interesting here to study the toy model at $[0,+\infty)$ and use the Dirichlet boundary conditions. Setting $\psi_{-2}(0)=0$ does not impose any restriction on the coefficients $A,B$, therefore allowing the outgoing wave to have infinite amplitude. This invariance is fixed through regularisation. Regularisation is a procedure aimed at handling the divergences of an integral. In the case of singular potentials, the integral of the squared modulus of the wave function diverges, and several methods can be employed to address this issue. One approach is to modify the measure by slightly altering the dimension of the differential element in the integral. Another is to adjust the integrand by manipulating the solution. Alternatively, we can modify the domain of integration. In this context, the simplest method is to introduce a dimensional parameter that breaks scale invariance and alters the domain of the Hamiltonian. But why does this work?

Let us first outline the mathematical essence of the problem and then proceed to provide a physical and intuitive solution. We know that observables correspond to self-adjoint operators. A self-adjoint operator is a Hermitian operator (i.e., \( A = A^{\dagger} \)) on a dense subset of \( L_2 \), where the domains of \( A \) and \( A^{\dagger} \) coincide. In the case at hand, the derivatives of the functions we are dealing with will often diverge at the origin, preventing the Hamiltonian from being a Hermitian operator. To address this, we impose the condition that the allowed functions vanish in a neighbourhood of zero (not just at \( z = 0 \)), thereby modifying the domain of the operator \( H \). This modification simultaneously alters the domain of \( H^{\dagger} \) such that the domains no longer coincide. Making them coincide requires a secondary manoeuvre, known as the self-adjoint extension of \( H \), which can be performed thanks to the work of Von Neumann \cite{araujo2004operator}.

Knowing this, we will set aside the concern of extending the Hamiltonian and instead outline the process of making the Hamiltonian a Hermitian operator through the use of a cut-off. Specifically, we will require that the solutions vanish at a distance \( L>0 \) from the origin
\begin{equation}
    \psi_{-2}(L)=0.
\end{equation}

A good idea is to define $\Gamma_{-2}=\mu^2+1/4$, as we are interested in $\Gamma_{-2}>1/4$. According to Ref.~\cite{gupta1993renormalization}, the bounded energy levels are now given by
\begin{equation}
    E_n=-e^{\frac{-2\pi n}{\mu}}\left(\frac{2}{Le^{\gamma}}\right)^2+O(\mu).
\end{equation}

At first glance, even when $n=1$, taking $L\rightarrow 0$ will bring the anomaly back. This does not make sense; observable quantities, like energy, should not depend on unphysical parameters like the cut-off. However, we can consider $\mu(L)$ such that when taking the limit, it allows a physical ground state to exist. This relationship between $\mu$ and $L$ is captured by the so-called beta function, a key concept in renormalisation group theory. It is defined as
\begin{equation}
    \beta(\mu)=-L\frac{d\mu}{dL}.
\end{equation}

Now, recall that the relation captures the independence of the ground state with the cut-off
\begin{equation}
    \frac{\md E_1}{\md L}=\frac{\partial E_1}{\partial L}+\frac{\partial E_1}{\partial\mu}\frac{d\mu}{dL}=0
\end{equation}
yielding
\begin{equation}
    \frac{\md E_1}{\md L} = \me^{\left(-\frac{2\pi}{\mu(L)}\right)} \cdot \frac{4}{L^3 \me^{2\gamma}} \cdot \left[-2 + \frac{\pi L}{\mu^2(L)} \cdot \frac{\md\mu}{\md L} \right] +O(\mu(L))=0.
\end{equation}
This gives
\begin{equation}\label{differential_mu}
    -L\frac{\md\mu}{\md L}\approx-\frac{2\mu^2}{\pi^2},
\end{equation}
therefore determining
\begin{equation}
    \beta(\mu)\approx-\frac{2\mu^2}{\pi^2}.
\end{equation}
On the other hand, solving the differential equation given by (\ref{differential_mu}), one finds
\begin{equation}
    \mu(L)=-\frac{\pi}{C+2\log(L)}\rightarrow0\ \ \ \ \text{when  }\ L\rightarrow0
\end{equation}
and therefore $\beta(\mu)\rightarrow0$ when $L\rightarrow0$. This corresponds to the fact that the coupling becomes weaker at smaller scales. A similar treatment can be done within scattering theory by fixing a phase shift, involving the usage of beta functions. This, however, leaves us with one well-defined ground state and no excited states.

However, when the indexes of the Bessel functions are imaginary, both functions in the combination behave essentially the same. The scale symmetry is broken by introducing a scaling parameter—in our case, the reflection point—which enables the incorporation of non-trivial boundary conditions. In summary, the very framework of quantum reflection, as demonstrated in the paper, introduces a scale parameter—the reflection point—that breaks scale invariance. However, we will see that certain features of this symmetry persist and manifest in the results we obtain. With this in mind, we can now proceed to determine the reflectance as expected.

\begin{eqnarray}\label{R-2}
    R_{-2}=\left|C_{-2}\cdot e^{\pi \sqrt{\Gamma_{-2}-1/4}}\right|^2\,,
\end{eqnarray}
with
\begin{eqnarray}\label{C-2}
    C_{-2}=\frac{H_1'(qz_1)+\left(\frac{1}{2qz_1}+\mi\frac{q_1}{q}\right)H_1(qz_1)}{H_2'(qz_1)+\left(\frac{1}{2qz_1}+\mi\frac{q_1}{q}\right)H_2(qz_1)}\,,
\end{eqnarray}
and $H_j=H^{(j)}_{\mi\sqrt{\Gamma_{-2}-1/4}}$.

Recall that the expression of the reflection point $z_1$ for the potential $n=-2$ can be easily manipulated to yield $z_1 = \frac{1}{2q}\sqrt{\Gamma_{-2}}$ so that $qz_1 = \frac{1}{2}\sqrt{\Gamma_{-2}}$ and that $\frac{\Gamma_{-2}}{q^2z_1^2}=4$. Plugging what we have just deduced, we can see that
\begin{eqnarray}
    C_{-2}=\frac{H\ '\ ^{(1)}_{\mi\sqrt{\Gamma_{-2}-1/4}}\left(\frac{1}{2}\sqrt{\Gamma_{-2}}\right)+\left[\frac{1}{\sqrt{\Gamma_{-2}}}+5\mi\right]H ^{(1)}_{\mi\sqrt{\Gamma_{-2}-1/4}}\left(\frac{1}{2}\sqrt{\Gamma_{-2}}\right)}{H\ '\ ^{(2)}_{\mi\sqrt{\Gamma_{-2}-1/4}}\left(\frac{1}{2}\sqrt{\Gamma_{-2}}\right)+\left[\frac{1}{\sqrt{\Gamma_{-2}}}+5\mi\right]H ^{(2)}_{\mi\sqrt{\Gamma_{-2}-1/4}}\left(\frac{1}{2}\sqrt{\Gamma_{-2}}\right)}
\end{eqnarray}
where $\Gamma_{-2}$ is the dimensionless parameter. Therefore, $R_{-2}$ does not depend on the energy, just as we have found before. Moreover, $R_{-2}$ erases all dependence on any scale parameter, being only a function of the dimensionless coupling $\Gamma_{-2}$. This is, once again, a manifestation of the scale invariance presented by the potential $n=-2$.

\subsection{Singular potentials}
For $n<-2$, one finds that the Hamiltonian is never again a Hermitian operator, and therefore there is an overpopulation of eigenstates. Despite needing regularisation, one can solve the Schrödinger equation and then worry about the physics.

\subsubsection{n=-3}
For $n=-3$, let us consider the Schrödinger equation given by
\begin{equation}
    \psi''(z)+\frac{\Gamma_{-3}}{z^3}\psi(z)+q^2\psi(z)=0
\end{equation}
Following Ref.~\cite{case1950singular}, consider the change of variables
\begin{equation}
    y=\frac{1}{\sqrt{z}} \,,
\end{equation}
and
\begin{equation}
    \psi(y)=\me^{\alpha y}\varphi(y)\,,
\end{equation}
where we choose $\alpha^2=-4\Gamma_{-3}$. Let us then consider the following differential equation
\begin{equation}
    \varphi''(y)+(2\alpha+\frac{3}{y})\varphi'(y)+\bigg(\frac{3\alpha}{y}+\frac{4q^2}{y^6}\bigg)\varphi(y)=0,\label{eq:n=-3ode}
\end{equation}
and let us suggest an ansatz of the form
\begin{eqnarray}
    \varphi(y)&=&y^r\sum_{n=0}^{\infty}c_ny^{-n}\,,\\
    \varphi'(y)&=&y^r\sum_{n=0}^{\infty}c_n(r-n)y^{-n-1}\,,\\
    \varphi''(y)&=&y^r\sum_{n=0}^{\infty}c_n(n-r)(n-r+1)y^{-n-2}\,.
\end{eqnarray}
Inserting these expressions into equation~\eqref{eq:n=-3ode} yields
\begin{equation}
    \begin{split}
        \sum_{n=0}^{\infty}(n-r)(n-r+1)c_ny^{-n-2}-2\alpha\sum_{n=0}^{\infty}(n-r)c_ny^{-n-1}-3\sum_{n=0}^{\infty}(n-r)c_ny^{-n-2}+\\
        +3\alpha\sum_{n=0}^{\infty}c_ny^{-n-1}-4\eta^2\sum_{n=0}^{\infty}c_ny^{-n-6}=0.
    \end{split}
\end{equation}
Let us look into the form of the first coefficients. For $y^{-1}$, we have
\begin{equation*}
    (2\alpha r+3\alpha)c_0=0\rightarrow c_0=0\ \ \text{or} \ \ r=-3/2.
\end{equation*}
We will take the arbitrary coefficient $c_0\neq0$ and therefore $r=-3/2$. For the next coefficient, we get
\begin{equation*}
    [-r(1-r)-3(-r)]c_0+[-2\alpha(1-r)+3\alpha]c_1=0\rightarrow c_1=-\frac{r(r+2)}{2\alpha(r+1/2)}c_0.
\end{equation*}
Similarly, for the first few coefficients and plugging $r=-3/2$ we get
\begin{equation}
    c_n=\frac{(n+1/2)(n-3/2)}{2\alpha n}c_{n-1}=\frac{(3/2)_n\cdot (-1/2)_n}{(2\alpha)^n\cdot n!}c_0,\ \ \ \ \ 0\leq n\leq 4
\end{equation}
where the \textit{Pochhammer symbol} is defined by
\begin{equation*}
     (z)_n=\frac{\Gamma(z+n)}{\Gamma(z)}=z\cdot(z+1)\cdot...\cdot(z+n-1).
\end{equation*}
For $n\geq 5$, the term with $4q^2$ starts to have an effect and the coefficients include
\begin{equation}
    c_n=\frac{(n+1/2)(n-3/2)}{2n\alpha }c_{n-1}+\frac{2q^2}{n\alpha }c_{n-5}\,.
\end{equation}

Recall that this solution is correct in the range of big $y$, that is, near the singularity of the actual potential. This series can be rearranged so that the powers are $(z/\Gamma_{-3})^{n/2}$ and the coefficients are dimensionless. Now, depending on the choice of $\alpha=\pm 2\mi\sqrt{\Gamma_{-3}}$, we find two solutions given by
\begin{eqnarray}
    c_n^+=-\mi\frac{(n+1/2)(n-3/2)}{4n}c^+_{n-1}-\mi\frac{q^2\Gamma_{-3}^2}{n}c^+_{n-5}\quad \text{and}\quad c_n^-=(-1)^nc_n^+\ 
\end{eqnarray}
and writing again $c_n^+=c_n$,
\begin{eqnarray}
    S_+=\sum_{n=0}^{\infty}c_n\left(\sqrt{\frac{z}{\Gamma_{-3}}}\right)^n,\quad\text{and}\quad S_-=\sum_{n=0}^{\infty}(-1)^nc_n\left(\sqrt{\frac{z}{\Gamma_{-3}}}\right)^n\ .
\end{eqnarray}
Then, a general solution can be found to be
\begin{eqnarray}
    \Psi_{-3}(z)=\left(\sqrt{\frac{z}{\Gamma_{-3}}}\right)^{3/2}\left[C_1e^{2\mi\sqrt{\Gamma_{-3}/z}}\sum_{n=0}^{\infty}c_n\left(\sqrt{\frac{z}{\Gamma_{-3}}}\right)^n+C_2e^{-2\mi\sqrt{\Gamma_{-3}/z}}\sum_{n=0}^{\infty}(-1)^nc_n\left(\sqrt{\frac{z}{\Gamma_{-3}}}\right)^n\right]\ .
\end{eqnarray}

Note that this agrees with the expectations in \cite{case1950singular}, for which
\begin{equation}
    \Psi_{-3} \underset{z \to 0}{\sim} A z^{3/4} \cos\left(2\sqrt{\frac{\Gamma_{-3}}{z}} + B\right)\ .
\end{equation}

Now, as said, this solution is only valid for small $z$. If we want to study the asymptotic behaviour, we might want to use the WKB approximation:
\begin{eqnarray}
    \psi_{WKB}(z) \approx \frac{C}{\sqrt{k(z)}} \exp\left(\pm i \int^z k(z') dz'\right).
\end{eqnarray}
For
\begin{eqnarray}
    \int k(z') dz' = \int \sqrt{q^2 + \frac{\Gamma_{-3}}{z'^3}} dz'
\end{eqnarray}
and using
\begin{eqnarray}
    \sqrt{q^2 + \frac{\Gamma_{-3}}{z'^3}} = q \left(1 + \frac{\Gamma_{-3}}{q^2 z'^3}\right)^{1/2} \approx q \left(1 + \frac{1}{2} \frac{\Gamma_{-3}}{q^2 z'^3}\right) = q + \frac{\Gamma_{-3}}{2q z'^3}
\end{eqnarray}
one can integrate
\begin{eqnarray}
    \int \left(q + \frac{\Gamma_{-3}}{2q z'^3}\right) dz' = q z' + \frac{\Gamma_{-3}}{2q} \int z'^{-3} dz'
= q z' + \frac{\Gamma_{-3}}{2q} \frac{z'^{-2}}{-2} + \text{constant}
= q z' - \frac{\Gamma_{-3}}{4q z'^2} + \text{constant}\ .
\end{eqnarray}

The amplitude factor reads
\begin{eqnarray}
    \frac{1}{\sqrt{k(z)}} \approx \frac{1}{\sqrt{q}} \left(1 + \frac{\Gamma_{-3}}{q^2 z^3}\right)^{-1/4} \approx \frac{1}{\sqrt{q}} \left(1 - \frac{\Gamma_{-3}}{4q^2 z^3}\right).
\end{eqnarray}

Therefore,
\begin{eqnarray}
    \psi_{WKB}(z) \approx \frac{C_+}{\sqrt{q}} \exp\left(\mi \left(q z - \frac{\Gamma_{-3}}{4q z^2} + \delta_+\right)\right) + \frac{C_-}{\sqrt{q}} \exp\left(-\mi \left(q z - \frac{\Gamma_{-3}}{4q z^2} + \delta_-\right)\right)
\end{eqnarray}
where $\delta_+$ and $\delta_-$ incorporate the integration constants. We can combine these into a single phase shift $\phi$ and an overall amplitude D:
\begin{eqnarray}
    \psi(z) \underset{z \to \infty}{\sim} \frac{D}{\sqrt{q}} \cos\left(q z - \frac{\Gamma_{-3}}{4q z^2} + \phi\right)\ .
\end{eqnarray}
Relating the constants requires solving the "connection problem" – finding how the solution behaves in the intermediate region between small and large z and matching the different asymptotic forms.

\subsubsection{n=-4}
For $n=-4$, let us consider
\begin{equation}
    \psi''(z)+q^2\psi(z)+\frac{\Gamma_{-4}}{z^4}\psi(x)=0\,,
\end{equation}
and suggest a solution with the form $\psi(z)=\sqrt{z}\cdot \varphi(z)$. Then $\psi'(z)=\frac{1}{2\sqrt{z}}\varphi(z)+\sqrt{z}\varphi'(z)$ and $\psi''(z)=-\frac{1}{4}z^{-3/2}\varphi(z)+z^{-1/2}\varphi'(z)+\sqrt{z}\varphi''(z)$. All together converts the previous equation into
\begin{equation}
    \varphi''(z)+\frac{1}{z}\varphi'(z)+\Big(q^2+\frac{\Gamma_{-4}}{z^4}-\frac{1}{4z^2}\Big)\varphi(z)=0\,.
\end{equation}
Next, we re-define $y=\lambda z$ for some $\lambda\in\mathbb{R}\backslash\{0\}$ that we will pick later. The previous equation now becomes
\begin{equation*}
    \varphi''(y)+\frac{1}{y}\varphi'(y)+\Big(\frac{q^2}{\lambda^2}-\frac{1}{4y^2}+\frac{\lambda^2\Gamma_{-4}}{y^4}\Big)\varphi(y)=0.
\end{equation*}
Following Ref.~\cite{frank1971singular}, our last manipulation will be to change $y=\me^{-x}$. Recall that $\md y/\md x=-\me^{-x}$ and therefore $\md/\md x=-\me^{-x}\md/\md y$ implies $\md/\md y=-\me^x\md/\md x$. Then $\md^2/\md y^2=-\me^{2x}(\md/\md x+\md^2/\md x^2)$. Now, choosing $\lambda=(q^2/\Gamma_{-4})^{1/4}$ and $h^2=q^2/\lambda^2$, we obtain the following equation
\begin{equation}
    \varphi''(x)-\Big(\frac{1}{4}-2h^2\cosh(2x)\Big)\varphi(x)=0\,,
\end{equation}
which is known as the modified Mathieu equation and has solutions, the modified Mathieu functions. That is,
\begin{equation}
    \psi_{-4}(z)=C_1\sqrt{z}M_{\nu}^{(3)}(\pm\ln(\lambda z),h)+C_2\sqrt{z}M_{\nu}^{(4)}(\pm\ln(\lambda z),h)\,,
\end{equation}
where the $+$ sign is taken if $|\lambda z|\geq1$ and the $-$ sign if $|\lambda x|\leq1$ (its derivative does not exist for $z=1/\lambda$). The parameter $\nu$, on the other hand, is a complicated function of $q$ and $a$. For details on modified Mathieu functions, we refer to Ref.~\cite{van2007accurate}. An expression for $\nu$ can be found in Ref.~\cite{spector1964exact}. We have
\begin{eqnarray*}
    \lefteqn{\cos \nu \pi = \cos \Gamma_{-4}^{\frac{1}{4}} \pi + \frac{\pi \sin \Gamma_{-4}^{\frac{1}{4}} \pi}{4\Gamma_{-4}^{\frac{3}{4}}(\Gamma_{-4} - 1)} q^2 +
\left[ \frac{15\Gamma_{-4}^2 - 35\Gamma_{-4} + 8}{64(\Gamma_{-4} - 1)^3 (\Gamma_{-4} - 4)} \pi \sin \Gamma_{-4}^{\frac{1}{4}} \pi 
- \frac{\pi^2 \cos \Gamma_{-4}^{\frac{1}{4}} \pi}{32\Gamma_{-4} (\Gamma_{-4} - 1)^2} \right] q^4}\\
&&+ \left[ \frac{105\Gamma_{-4}^5 - 1155\Gamma_{-4}^4 + 3815\Gamma_{-4}^3 - 4705\Gamma_{-4}^2 + 1652\Gamma_{-4} - 288}{256(\Gamma_{-4} - 1)^6 (\Gamma_{-4} - 4)^2 (\Gamma_{-4} - 9) \Gamma_{-4}^{5/2}} \pi \sin \Gamma_{-4}^{\frac{1}{4}} \pi\right.\\
&&\left.
- \frac{\pi^3 \sin \Gamma_{-4}^{\frac{1}{4}} \pi}{384(\Gamma_{-4} - 1)^3}
- \frac{15\Gamma_{-4}^2 - 35\Gamma_{-4} + 8}{256\Gamma_{-4}^2 (\Gamma_{-4} - 1)^4 (\Gamma_{-4} - 4)} \pi^2 \cos \Gamma_{-4}^{\frac{1}{4}} \pi \right] q^6 + \dots
\end{eqnarray*}

Alternatively, this can again be solved with a series approach. One finds
\begin{eqnarray}
    \Psi_{-4}=z\left[C_1\me^{\mi\sqrt{\Gamma_{-4}}/z}-C_2\me^{-\mi\sqrt{\Gamma_{-4}}/z}\right]\sum_{n=0}^{\infty}c_nz^n
\end{eqnarray}
for
\begin{eqnarray}
    c_n=-\mi\frac{n-1}{2\sqrt{\Gamma_{-4}}}c_{n-1}-\mi\frac{q^2}{2n\sqrt{\Gamma_{-4}}}c_{n-3}.
\end{eqnarray}
One can see that
\begin{eqnarray*}
    \frac{\md\Psi_{-4}}{\md z} = \left[C_1 \me^{\mi\sqrt{\Gamma_{-4}}/z} - C_2 \me^{-\mi\sqrt{\Gamma_{-4}}/z}\right] \sum_{n=0}^\infty c_n (1 + n) z^n - \frac{\mi\sqrt{\Gamma_{-4}}}{z} \left[C_1 \me^{\mi\sqrt{\Gamma_{-4}}/z} + C_2 \me^{-\mi\sqrt{\Gamma_{-4}}/z}\right] \sum_{n=0}^\infty c_n z^n\,.
\end{eqnarray*}

Just as before, this solution is valid only near the origin.

\section{Reflection at a continuous potential}
\subsection{The Badlands function}
The reflection occurs at the maximum of the badlands function
\begin{equation}
    \left|B(z)\right| = \left|\hbar^2\left(\frac{3}{4}\frac{\left[p'(z)\right]^2}{p^4(z)} - \frac{1}{2}\frac{p''(z)}{p^3(z)}\right)\right| \ll 1 \,.\label{eq:badlands}
\end{equation}
Using the potential $U(z) = \lambda z^n +U_0$, a classical returning point can be evaluated by balancing the kinetic and potential energy
\begin{equation}
    E = U(z_{\rm c}) = \lambda z_{\rm c}^n + U_0 \,.
\end{equation}
Thus, the constant potential shift can be substituted, and the rescaled potential reads
\begin{eqnarray}
    U(z) = \lambda \left(z^n-z_{\rm c}^n\right) +E\,.
\end{eqnarray}
Consequently, the local classical momentum reads
\begin{equation}
    p(z) = \sqrt{2m \lambda \left(z_{\rm c}^n-z^n\right)} \,.
\end{equation}
One finds
\begin{eqnarray}
    |B(z)|=\left|\frac{\hbar^2nm^2\lambda^2z^{n-2}\left[\left(\frac{n}{4}+1\right)z^n+(n-1)z_c^n\right]}{p(z)^6}\right|\,,
\end{eqnarray}
leading to
\begin{eqnarray}\label{derivative_bandlands}
    \frac{\md B}{\md z} &=& \frac{\hbar^2 n z^{n-3}}{8 m \lambda (z_c^n - z^n)^4} \left[ \frac{(n+4)(n+2)}{4} z^{2n} + \right. \\
    && \left. \frac{(n-1)(5n+8)}{2} z^n z_c^n + (n-1)(n-2) z_c^{2n} \right]
\end{eqnarray}
and the maximum can be found by writing $t=z^n$ and solving a second-order polynomial, obtaining
\begin{eqnarray}
    z^n = z_c^n \frac{ -(n-1)(5n+8) \pm n\sqrt{3(n-1)(7n+13)} }{(n+4)(n+2)}\,,
\end{eqnarray}
where
\begin{eqnarray}
    z_c^n = \frac{E_{\perp}-U_0}{\lambda}\,,
\end{eqnarray}
which can be written as
\begin{equation}\label{reflection_point}
    z_0 = \sqrt[n]{\frac{-5n^2+\left(\sqrt{21n^2 +18n-39}-3\right)n+8}{n^2+6n+8}\frac{E_\perp-U_0}{\lambda}}\,.
\end{equation}

For $n=-2,-4$, one sees that the denominator diverges. This is, of course, because the quadratic term in Eq. (\ref{derivative_bandlands}) becomes zero. In these cases, the maximum is found to be in
\begin{eqnarray}
    z_0=\left|\frac{n-2}{\frac{5}{2}n-4}\right|^{1/n}z_c\ .
\end{eqnarray}

\subsection{Ricatti differential equation}
The Riccati differential equation reads
\begin{equation}
    \tilde{r}'(z) = 2\mi q(z)r(z) +\frac{q'(z)}{2q(z)}\left[1-r^2(z)\right]\,, \label{eq:ricatti2}
\end{equation}
with 
\begin{equation}
    q(z) = \sqrt{\frac{m^2v_z^2}{\hbar^2}-\frac{2m}{\hbar^2}U(z)}\,.
\end{equation}

The Ricatti differential equation can be solved analytically for the linear potential, leading to the solution
\begin{eqnarray}\label{n1_johannes}
    r_1=\frac{-I_{-1/3}(\tilde{x})+I_{2/3}(\tilde{x})-I_{-2/3}(\tilde{x}) +I_{1/3}(\tilde{x}) }{-I_{-1/3}(\tilde{x})-I_{2/3}(\tilde{x}) +I_{-2/3}(\tilde{x})+I_{1/3}(\tilde{x}) }\,,
\end{eqnarray}
with $\tilde{x}=\frac{\mathrm{i} m^{2} v_z^{3}}{3\hbar\tilde{\Gamma}_1}$.

\subsection{Numerical integration of the Ricatti differential equation}
The Runge--Kutta method in a forward scheme is applicable to solve this differential equation~(\ref{eq:ricatti2}). However, a one-dimensional potential is commonly defined in the positive half-space, and the interface is placed at the origin $z=0$. Thus, we switch the potential, $U(z) \mapsto U(-z)$ and shift the interface by a distance $\tilde{z}$, mathematically $U(z) \mapsto U(-z-\tilde{z})$. To apply the Runge--Kutta scheme to solve the differential equation, we initialise the solution without any reflection at the origin $r(0)=0$. This condition is ensured by choosing the shift to be one hundred times the distance of the reflection point $\tilde{z} = 100 z_0$, and thus, the simulation terminates when reaching this point.

The local momentum reads
\begin{equation}
    p(z) = \sqrt{2m\left[E-U(z)\right]} \,,
\end{equation}
and thus its derivatives read
\begin{eqnarray}
    p'(z) &=& -\frac{m}{p(z)}U'(z)\,,\\
    p''(z)&=& -\frac{m^2}{p^3(z)}\left[U'(z)\right]^2 -\frac{m}{p(z)}U''(z)\,,\\
    p'''(z) &=& -\frac{3 m^3}{p^5(z)}\left[U'(z)\right]^3 - \frac{3m^2}{p^3(z)}U'(z)U''(z)-\frac{m}{p(z)}U'''(z) \,.
\end{eqnarray}
Analogously, we find the derivative of $q$
\begin{eqnarray}
    q'(z) &=& -\frac{m}{\hbar^2 q(z)}U'(z) \,,\\
    \frac{q'(z)}{q(z)} &=& -\frac{m}{\hbar^2 q^2(z)}U'(z) \,.
\end{eqnarray}
\begin{eqnarray}
    \frac{\mathrm d}{\mathrm d z}\left|B(z)\right| = \frac{1}{\left|B(z)\right|} \left[ \frac{3}{4}\frac{p'(z)}{p^4(z)} -\frac{1}{2}\frac{p''(z)}{p^3(z)}  \right]\times \left[ \frac{3}{4}\frac{p''(z)}{p^4(z)}-3\frac{\left[p'(z)\right]^2}{p^5(z)}-\frac{1}{2}\frac{p'''(z)}{p^3(z)}+\frac{3}{2}\frac{p''(z)p'(z)}{p^4(z)}\right]
\end{eqnarray}

\section{Interference patterns}
The interference pattern at a regular grating is given by
\begin{eqnarray}
I(x) =  \frac{a_0^2K^2}{4\pi^2L_1^2L_2^2}\left|\sum_{n=1}^N\int\limits_{pn-w/2}^{pn+w/2} \mathrm d s\mathrm e^{-\mi Ks\frac{x}{L_2}} f(s)\right|^2\,,
\end{eqnarray}
with the amplitude $a_0$, the wave vector $K=mv_z/\hbar$, the distance from the source to the grating $L_1$, the distance from the grating to the screen $L_2$, the period $p$, the number of slits $N$, the opening width $w$, and  function $f(s)$, which describes the phase shift due to the Casimir--Polder interaction (we neglect effects of rotation on the phase) for the transmitted wave
\begin{equation}
f(s) = \mathrm e^{-\mi \frac{C_3d}{\hbar v_z}\left(\frac{1}{(w/2-s)^3} +\frac{1}{\left(s+w/2\right)^3}\right)}\,,  
\end{equation}
and for the reflected wave
\begin{eqnarray}
    f(s) = R(\vartheta,K) \me^{2\mi\sin\vartheta} \,.
\end{eqnarray}
The period of the grating changes by rotation $p=p_0\cos\vartheta$. Analogously, the width of the opening splits into a region where the wave is transmitted and a region reflecting the wave. The length of the transmission region can be determined by
\begin{equation}
    w_{\rm T} = p_0(1-c) \cos\vartheta -d\sin\vartheta\,,
\end{equation}
where we introduced the opening fraction $c$. The second term describes the area where the wave will be reflected. Thus, the corresponding opening width reads
\begin{equation}
    w_{\rm R} = d\sin\vartheta\,.
\end{equation}
Hence, the interference pattern for the reflected beam reads
\begin{equation}
    I_{\rm R} = I_0R^2(\vartheta,K)\frac{\sin^2\left[\frac{N}{2}Kp_0\cos\vartheta\left(\tan\alpha-2\sin\vartheta\right)\right]}{\sin^2\left[\frac{Kp_0\cos\vartheta}{2}\left(\tan\alpha-2\sin\vartheta\right)\right]}\frac{\sin^2\left[\frac{K d\sin\vartheta}{2}\left(\tan\alpha-2\sin\vartheta\right)\right]}{\left(\tan\alpha-2\sin\vartheta\right)^2}\,,
\end{equation}
with $\tan\alpha = x/L_2$ and $I_0 = \frac{a_0^2}{\pi^2L_1^2L_2^2}$. Analogously, one finds
\begin{equation}
    I_{\rm T} = \frac{I_0 K^2}{4}\frac{\sin^2\left[\frac{N}{2}Kp_0\cos\vartheta\tan\alpha\right]}{\sin^2\left[\frac{Kp_0\cos\vartheta}{2}\tan\alpha\right]}\left|\int\limits_{-w_{\rm T}/2}^{w_{\rm T}/2}\md s \,\me^{\mi K s \tan\alpha }\me^{-\mi \frac{C_3d}{\hbar v_z}\left[\frac{1}{\left(w_{\rm T}/2-s\right)^3}+\frac{1}{\left(s+w_{\rm T}/2\right)^3}\right]}\right|^2\,,
\end{equation}
which cannot be carried out analytically.

\bibliographystyle{unsrt}  
\bibliography{bibi.bib}

\begin{thebibliography}{10}

\bibitem{Brenig1980}
W~Brenig.
\newblock On the low energy limit of reflection and sticking coefficients in atom surface scattering. i. short range forces.
\newblock {\em Z. Phys. B}, 26:227--233, 1980.

\bibitem{Boeheim1982}
J.~B\"oheim et~al.
\newblock On the low energy limit of reflection and sticking coefficients in atom surface scattering.
\newblock {\em Z. Phys. B}, 48:43--49, 1982.

\bibitem{PhysRevA.65.032902}
H.~Friedrich et~al.
\newblock Quantum reflection by casimir--van der waals potential tails.
\newblock {\em Phys. Rev. A}, 65:032902, Feb 2002.

\bibitem{PhysRevLett.91.193202}
V.~Druzhinina and M.~DeKieviet.
\newblock Experimental observation of quantum reflection far from threshold.
\newblock {\em Phys. Rev. Lett.}, 91:193202, Nov 2003.

\bibitem{10.1063/1.3246162}
H.~Khemliche et~al.
\newblock Grazing incidence fast atom diffraction: An innovative approach to surface structure analysis.
\newblock {\em Appl. Phys. Lett.}, 95(15):151901, 10 2009.

\bibitem{PhysRevLett.97.093201}
T.~A.~Pasquini et~al.
\newblock Low velocity quantum reflection of bose-einstein condensates.
\newblock {\em Phys. Rev. Lett.}, 97:093201, Aug 2006.

\bibitem{PhysRevA.71.052901}
H.~Oberst et~al.
\newblock Quantum reflection of ${\mathrm{he}}^{*}$ on silicon.
\newblock {\em Phys. Rev. A}, 71:052901, May 2005.

\bibitem{Buhmann12a}
S.~Y. Buhmann.
\newblock {\em {Dispersion Forces I: Macroscopic quantum electrodynamics and ground-state Casimir, Casimir--Polder and van der Waals forces}}.
\newblock Springer, Heidelberg, 2012.

\bibitem{Buhmann12b}
S.~Y. Buhmann.
\newblock {\em {Dispersion Forces II: Many-Body Effects, Excited Atoms, Finite Temperature and Quantum Friction}}.
\newblock Springer Tracts in Modern Physics. Springer, Heidelberg, 2012.

\bibitem{D2CP03349F}
J.~Fiedler et~al.
\newblock Perspectives on weak interactions in complex materials at different length scales.
\newblock {\em Phys. Chem. Chem. Phys.}, 25:2671--2705, 2023.

\bibitem{RevModPhys.81.1051}
A.D.~Cronin et~al.
\newblock Optics and interferometry with atoms and molecules.
\newblock {\em Rev. Mod. Phys.}, 81:1051--1129, Jul 2009.

\bibitem{PhysRevA.108.023306}
J.~Fiedler et~al.
\newblock Monolithic atom interferometry.
\newblock {\em Phys. Rev. A}, 108:023306, Aug 2023.

\bibitem{Fiedler2024}
J.~Fiedler and B.~Holst.
\newblock A continuous beam monochromator for matter waves.
\newblock {\em Eur. Phys. J. D}, 78:39, 2024.

\bibitem{Oberst2005}
H.~Oberst and F.~Shimizu.
\newblock Quantum reflection of cold atoms.
\newblock {\em J. Phys.: Conf. Ser.}, 19:158--165, 2005.

\bibitem{Ite1993}
I.~A.~Yu et~al.
\newblock Evidence for universal quantum reflection of hydrogen from liquid 4he.
\newblock {\em Phys. Rev. Lett.}, 71, 1993.

\bibitem{Grucker2007}
J.~Grucker et~al.
\newblock Diffraction of fast metastable atoms by micrometric reflection gratings.
\newblock {\em Eur. Phys. J. D}, 41:467–474, 2007.

\bibitem{PhysRevLett.86.987}
F.~Shimizu.
\newblock Specular reflection of very slow metastable neon atoms from a solid surface.
\newblock {\em Phys. Rev. Lett.}, 86:987--990, Feb 2001.

\bibitem{Zhao2008}
B.~Suk~Zhao et~al.
\newblock Quantum reflection of helium atom beams from a microstructured grating.
\newblock {\em Phys. Rev. A}, 78, 2008.

\bibitem{Pasquini2004}
T.~A.~Pasquini et~al.
\newblock Quantum reflection of helium atom beams from a microstructured grating.
\newblock {\em Phys. Rev. Lett.}, 93, 2004.

\bibitem{Pasquini2006}
T.~A.~Pasquini et~al.
\newblock Low velocity quantum reflection of bose-einstein condensates.
\newblock {\em Phys. Rev. Lett.}, 97, 2006.

\bibitem{Marchant2016}
A.~L.~Marchant et~al.
\newblock Quantum reflection of bright solitary matter waves from a narrow attractive potential.
\newblock {\em Phys. Rev. A}, 93, 2016.

\bibitem{Dufour2014}
G.~Dufour et~al.
\newblock Quantum reflection of bright solitary matter waves from a narrow attractive potential.
\newblock {\em Int. J. Mod. Phys. Conf. Ser.}, 30, 2014.

\bibitem{Bai2020}
Z.~Bai et~al.
\newblock Quantum reflections of nonlocal optical solitons in a cold rydberg atomic gas.
\newblock {\em Phys. Rev. A}, 101, 2020.

\bibitem{Bai2019}
Y.-P.~Bai et~al.
\newblock Model for investigating quantum reflection and quantum coherence in ultracold molecular collisions.
\newblock {\em Phys. Rev. A}, 101, 2019.

\bibitem{Rojas2020}
G.~Rojas-Lorenzo et~al.
\newblock Quantum threshold reflection of he-atom beams from rough surfaces.
\newblock {\em Phys. Rev. A}, 101, 2020.

\bibitem{Kilianski2024}
R.~Kilianski and R.~Bennett.
\newblock Designer quantum reflection from a micropore.
\newblock {\em Phys. Rev. A}, 109, 2024.

\bibitem{Lecoffre2025}
J.~Lecoffre et~al.
\newblock Measurement of casimir-polder interaction for slow atoms through a material grating.
\newblock {\em Phys. Rev. Research}, 7, 2025.

\bibitem{Lekner2016}
J.~Lekner.
\newblock {\em {Theory of Reflection}}, volume~87 of {\em Springer Series on Atomic, Optical, and Plasma Physics}.
\newblock Springer International Publishing, Cham, 2016.

\bibitem{Born1999}
M.~Born and E.~Wolf.
\newblock {\em {Principles of Optics}}.
\newblock Cambridge University Press, oct 1999.

\bibitem{jackson1998classical}
J.D. Jackson.
\newblock {\em Classical Electrodynamics}.
\newblock Wiley, New York, 1998.

\bibitem{Chew}
W.~C. Chew.
\newblock {\em Waves and Fields in Inhomogenous Media}.
\newblock IEEE Press, 1995.

\bibitem{c9030064}
G.~L. Klimchitskaya and V.~M. Mostepanenko.
\newblock Casimir–polder force on atoms or nanoparticles from gapped and doped graphene: Asymptotic behavior at large separations.
\newblock {\em C}, 9(3), 2023.

\bibitem{doi:10.1126/sciadv.1500901}
B.~S.~Zhao et~al.
\newblock Universal diffraction of atoms and molecules from a quantum reflection grating.
\newblock {\em Sci. Adv.}, 2(3):e1500901, 2016.

\bibitem{PhysRevA.94.012513}
N.~Khusnutdinov et~al.
\newblock Casimir-polder effect for a stack of conductive planes.
\newblock {\em Phys. Rev. A}, 94:012513, Jul 2016.

\bibitem{Doak_2000}
R.B. Doak and A.V.G. Chizmeshya.
\newblock Sufficiency conditions for quantum reflection.
\newblock {\em EPL}, 51(4):381, 2000.

\bibitem{PhysRevB.64.085418}
A.~Mody et~al.
\newblock No-sticking effect and quantum reflection in ultracold collisions.
\newblock {\em Phys. Rev. B}, 64:085418, 2001.

\bibitem{Scheel2008}
S.~Scheel and S.~Y. Buhmann.
\newblock {Macroscopic QED - concepts and applications}.
\newblock {\em Acta Physica Slovaca}, 58(5):675--809, feb 2008.

\bibitem{PhysRevB.101.235424}
J.~Fiedler et~al.
\newblock Nontrivial retardation effects in dispersion forces: From anomalous distance dependence to novel traps.
\newblock {\em Phys. Rev. B}, 101:235424, Jun 2020.

\bibitem{Osestad2025}
E.~K.~Osestad et~al.
\newblock A novel gas sensing principle based on quantum fluctuations.
\newblock {\em EPJ Quantum Techn.}, 12:37, 2025.

\bibitem{Schollkopf1994}
W.~Sch\"ollkopf and J.~P. Toennies.
\newblock Nondestructive {M}ass selection of {S}mall {V}an der {W}aals {C}lusters.
\newblock {\em Science}, 266:1345--1348, 1994.

\bibitem{https://doi.org/10.1002/andp.201500214}
C.~Brand et~al.
\newblock A green's function approach to modeling molecular diffraction in the limit of ultra-thin gratings.
\newblock {\em Ann. Phys.}, 527(9-10):580--591, 2015.

\bibitem{Hemmerich16}
J.L.~Hemmerich et~al.
\newblock Impact of casimir-polder interaction on poisson-spot diffraction at a dielectric sphere.
\newblock {\em Phys. Rev. A}, 94:023621, Aug 2016.

\bibitem{Pars}
V.~A. Parsegian.
\newblock {\em Van der Waals forces: A handbook for biologists, chemists, engineers, and physicists}.
\newblock Cambridge University Press, New York, 2006.

\bibitem{Fiedler15}
J.~Fiedler and S.~Scheel.
\newblock {Casimir--Polder potentials on extended molecules}.
\newblock {\em Ann. Phys.}, 527(9-10):570--579, 2015.

\bibitem{C9CP03165K}
J.~Fiedler et~al.
\newblock Impact of effective polarisability models on the near-field interaction of dissolved greenhouse gases at ice and air interfaces.
\newblock {\em Phys. Chem. Chem. Phys.}, 21:21296--21304, 2019.

\bibitem{Fiedler_2022}
J.~Fiedler and B.~Holst.
\newblock An atom passing through a hole in a dielectric membrane: impact of dispersion forces on mask-based matter-wave lithography.
\newblock {\em J. Phys. B: At. Mol. Opt. Phys.}, 55(2):025401, feb 2022.

\bibitem{PhysRevA.85.042513}
S.Y.~Buhmann et~al.
\newblock Casimir-polder interaction of fullerene molecules with surfaces.
\newblock {\em Phys. Rev. A}, 85:042513, Apr 2012.

\bibitem{PhysRevResearch.6.023165}
C.~Garcion et~al.
\newblock Quantum description of atomic diffraction by material nanostructures.
\newblock {\em Phys. Rev. Res.}, 6:023165, May 2024.

\bibitem{Arndt1999}
M.~Arndt et~al.
\newblock Wave-particle duality of {C}60 molecules.
\newblock {\em Nature}, 401(6754):680--682, 1999.

\bibitem{Reisinger11}
T.~Reisinger et~al.
\newblock Particle–wave discrimination in poisson spot experiments.
\newblock {\em New J. Phys.}, 13(6):065016, 2011.

\bibitem{PhysRevLett.125.050401}
N.~Gack et~al.
\newblock Signature of short-range van der waals forces observed in poisson spot diffraction with indium atoms.
\newblock {\em Phys. Rev. Lett.}, 125:050401, Jul 2020.

\bibitem{Fiedler_2023}
J.~Fiedler et~al.
\newblock Realistic mask generation for matter-wave lithography via machine learning.
\newblock {\em Mach. learn.: sci. technol.}, 4(2):025028, jun 2023.

\bibitem{PhysRevA.109.032824}
L.~Queiroz.
\newblock Influence of retardation and dispersive surfaces on the regimes of the lateral casimir-polder force.
\newblock {\em Phys. Rev. A}, 109:032824, Mar 2024.

\bibitem{PhysRevA.97.023806}
J.C. de~A. Carvalho~et al.
\newblock Retardation effects in spectroscopic measurements of the casimir-polder interaction.
\newblock {\em Phys. Rev. A}, 97:023806, Feb 2018.

\bibitem{Hu:2004zu}
B.~L.~Hu et~al.
\newblock {Vacuum fluctuations and moving atoms / detectors: From Casimir-Polder to Unruh effect}.
\newblock {\em J. Opt. B}, 6:S698--S705, 2004.

\bibitem{ovdat2019breaking}
O.~Ovdat and E.~Akkermans.
\newblock {The breaking of continuous scale invariance to discrete scale invariance: a universal quantum phase transition}.
\newblock 9 2019.

\bibitem{brattan2018landscape}
D.K.~Brattan et~al.
\newblock On the landscape of scale invariance in quantum mechanics.
\newblock {\em J. Phys. A: Math. Theor.}, 51(43):435401, 2018.

\bibitem{camblong2001dimensional}
H.E.~Camblong et~al.
\newblock Dimensional transmutation and dimensional regularization in quantum mechanics: I. general theory.
\newblock {\em Ann. Phys.}, 287(1):14--56, 2001.

\bibitem{essin2006quantum}
A.~M. Essin and D.~J. Griffiths.
\newblock Quantum mechanics of the 1/ x2 potential.
\newblock {\em Am. J. Phys.}, 74(2):109--117, 2006.

\bibitem{gupta1993renormalization}
K.~S. Gupta and S.~G. Rajeev.
\newblock Renormalization in quantum mechanics.
\newblock {\em Phys. Rev. D}, 48(12):5940, 1993.

\bibitem{case1950singular}
K.M. Case.
\newblock Singular potentials.
\newblock {\em Phys. Rev.}, 80(5):797, 1950.

\bibitem{gradshteyn1988tables}
I.~S. Gradshteyn, I.~M. Ryzhik, and R.~H. Romer.
\newblock Tables of integrals, series, and products, 1988.

\bibitem{araujo2004operator}
V.S.~Araujo et~al.
\newblock Operator domains and self-adjoint operators.
\newblock {\em Am. J. Phys.}, 72(2):203--213, 2004.

\bibitem{frank1971singular}
W.~M.~Frank et~al.
\newblock Singular potentials.
\newblock {\em Rev. Mod. Phys.}, 43(1):36, 1971.

\bibitem{van2007accurate}
A.~Van~Buren and J.~Boisvert.
\newblock Accurate calculation of the modified mathieu functions of integer order.
\newblock {\em Quart. Appl. Math.}, 65(1):1--23, 2007.

\bibitem{spector1964exact}
R.~M. Spector.
\newblock Exact solution of the schr{\"o}dinger equation for inverse fourth-power potential.
\newblock {\em J. Math. Phys.}, 5(9):1185--1189, 1964.

\end{thebibliography}

\end{document}